\documentclass[10pt,journal,compsoc]{IEEEtran}
\newif\ifpeerreview

\peerreviewfalse
\usepackage[nocompress]{cite}
\usepackage{times}
\usepackage{soul}
\usepackage{url}
\usepackage[hidelinks]{hyperref}
\usepackage[utf8]{inputenc}
\usepackage[small]{caption}
\usepackage{graphicx}
\usepackage{amsmath, amssymb}
\usepackage{amsthm}
\usepackage{booktabs}
\usepackage{algorithm}
\usepackage{algorithmic}
\usepackage[switch]{lineno}
\usepackage{multirow}
\usepackage{graphicx}
\usepackage{colortbl}
\usepackage{enumitem}
\usepackage{float}
\usepackage{algorithm}
\usepackage{xcolor}
\usepackage{colortbl}
\usepackage{graphicx}
\usepackage{wrapfig} 
\usepackage{booktabs}

\usepackage[symbol]{footmisc}
\usepackage{tikz}
\usepackage{comment}
\usepackage{amsmath,amssymb} % define this before the line numbering.
\usepackage{color}
\usepackage[accsupp]{axessibility}  % Improves PDF readability for those with disabilities.
\usepackage{multicol}
\usepackage{lipsum}
\usepackage{orcidlink}

\title{PhotonSplat: 3D Scene Reconstruction and Colorization from SPAD Sensors}

\author{Sai Sri Teja*, Sreevidya Chintalapati*, Vinayak Gupta*, Mukund Varma T, Haejoon Lee, Aswin                    Sankaranarayanan, and Kaushik Mitra
\IEEEcompsocitemizethanks{\IEEEcompsocthanksitem Sai Sri Teja, Sreevidya Chintalapati, Vinayak Gupta and Kaushik Mitra are with the Department
of Electrical Engineering, Indian Institute of Technology, Madras, India
\IEEEcompsocthanksitem Mukund Varma T is with the Department of Computer Science, University of California, San Diego, CA 92093
\IEEEcompsocthanksitem Haejoon Lee and Aswin Sankaranarayanan are with the Department of Electrical and Computer Engineering, Carnagie Mellon University, Pittsburgh, PA, 15213
\IEEEcompsocthanksitem E-mail: \{saisriteja, sreevidya.chintalapati\}@gmail.com
\IEEEcompsocthanksitem * denotes equal contribution.}% <-this % stops an unwanted space
}

\begin{document}

\IEEEtitleabstractindextext{%
% \begin{abstract}
% The abstract goes here. \lipsum[1]
\begin{abstract}
% Advances in 3D reconstruction have enabled high-quality 3D capture possible, but requires a user to collect numerous images of a scene. Capturing clean, unblurry images of a scene is challenging, requiring limited motion of objects present in the scene and stable camera viewpoints. Recent advancements in camera technology like Single-Photon Avalanche Diode (SPAD) arrays allow image capture at extremely high frame rates, upto 100,000 frames per second. However, SPAD images are binary and heavily influenced by photon noise, which hinders their use in 3D reconstruction. In this paper, we present \textit{PhotonSplat}, a novel 3D reconstruction technique from multi-view SPAD image captures. To learn scene color, our framework uses generative priors or a single blurry color image of the scene - enabling further downstream tasks, including object recognition, and scene editing. We establish a new dataset for novel-view synthesis from SPAD images, dubbed \textit{PhotonScenes}, a real-world multi-view dataset captured using a SPAD sensor. All the code and data will be made publicly available upon acceptance. 

Advances in 3D reconstruction using neural rendering have enabled high-quality 3D capture. 
However, they often fail when the input imagery is corrupted by motion blur, due to fast motion of the camera or the objects in the scene. 
This work advances neural rendering techniques in such scenarios by using single-photon avalanche diode (SPAD) arrays, an emerging sensing technology capable of sensing images at extremely high speeds. However, the use of SPADs presents its own set of unique challenges in the form of binary images, that are driven by stochastic photon arrivals. To address this, we introduce \textit{PhotonSplat}, a framework designed to reconstruct 3D scenes directly from SPAD binary images, effectively navigating the noise vs.\ blur trade-off. Our approach incorporates a novel 3D spatial filtering technique to reduce noise in the renderings. The framework also supports both no-reference using generative priors and reference-based colorization from a single blurry image, enabling downstream applications such as segmentation, object detection and appearance editing tasks. Additionally, we extend our method to incorporate dynamic scene representations, making it suitable for scenes with moving objects. We further contribute \textit{PhotonScenes}, a real-world multi-view dataset captured with the SPAD sensors. Code, data and video results are available at \href{https://vinayak-vg.github.io/PhotonSplat/}{vinayak-vg.github.io/PhotonSplat/}.
\end{abstract} 
% \end{abstract}

}

% Make Title
% \ifpeerreview
% \linenumbers \linenumbersep 15pt\relax 
% \author{Paper ID \paperID\IEEEcompsocitemizethanks{\IEEEcompsocthanksitem This paper is under review for ICCP 2025 and the PAMI special issue on computational photography. Do not distribute.}}
% \markboth{Anonymous ICCP 2025 submission ID \paperID}%
% {}
% \fi
\maketitle

\begin{figure*}[h!]
\vspace{-2mm}
  \includegraphics[width=0.99\textwidth]{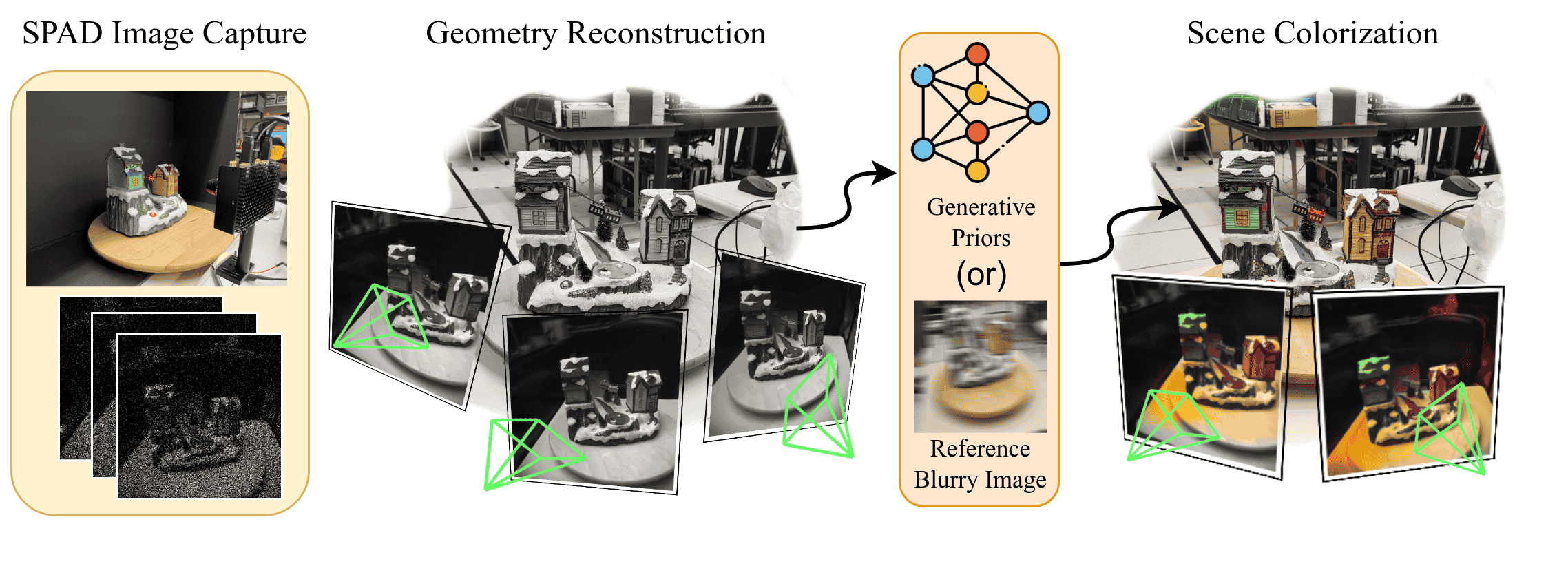}
\caption{In scenarios involving fast camera motion, such as drone surveillance, conventional RGB captures often suffer from severe motion blur, impeding accurate 3D structure modeling. To overcome this limitation, we employ Single-Photon Avalanche Diode (SPAD) camera arrays, which capture images at exceptionally high frame rates without motion blur. Our approach can successfully recover the underlying scene geometry from multi-view binary SPAD image captures. Furthermore, we demonstrate view-consistent colorization using either generative priors or a single blurry RGB image.
% Our objective is to reconstruct 3D scenes and render novel views from binary images. The GS model estimates photon-hitting probabilities by rendering $\lambda = \phi * \tau$ and applying the photon modeling equation. We optimize this using a binary cross-entropy loss between the rendered probabilities and the binary target image. To reduce overfitting to photon noise, we introduce a perturbation loss, comparing the rendered image with averaged outputs from perturbed camera views, ensuring smoother results. Additionally, a network learns transformations for the 3D Gaussians, rendering and averaging transformed views, which are supervised using ground truth blurry images to jointly deblur and colorize the scene.
}
\vspace{-2mm}
\label{fig:teaser}
\end{figure*}

% \twocolumn[{%
% \renewcommand\twocolumn[1][]{#1}%
% \maketitle\vspace{-2em}
% \includegraphics[width=.99\textwidth]{figures/teaserv4.png}
% \captionof{figure}{In scenarios involving fast camera motion, such as drone surveillance, conventional RGB captures often suffer from severe motion blur, impeding accurate 3D structure modeling. To overcome this limitation, we employ Single-Photon Avalanche Diode (SPAD) camera arrays, which capture images at exceptionally high frame rates without motion blur. Our approach can successfully recover the underlying scene geometry from multi-view binary SPAD image captures. Furthermore, we demonstrate view-consistent colorization using either generative priors or a single blurry RGB image.
% \vspace{1em}}
% \label{fig:teaser}
% }]

% \input{section/0_abstract} 
\vspace{-5mm}
\section{Introduction}

Recent advancements in photogrammetry techniques, driven by techniques like NeRF~\cite{mildenhall2021nerf} and Gaussian Splatting~\cite{kerbl3Dgaussians} have greatly improved accessibility to 3D scene reconstruction from 2D images. 
These methods require multi-view image captures of a real scene. They optimize a representation that captures the underlying 3D scene geometry and visual appearance, which then can be used to render images from any given viewpoint. 
Capturing images is, however, still challenging, requiring images that are  captured in well-lit conditions, free from motion blur. 
This is often challenging to achieve in practical settings, especially under low-light conditions, or with large object and camera motion. Thus, it is highly desirable to speed up the capturing process, where thousands of images could be captured in a fraction of a second. 

Traditional cameras, such as those utilizing CMOS sensors, accumulate a significant number of photons over a specified exposure time to generate high-quality, detailed images.
In specific scenarios like low-light environments, the exposure time needs to be increased to collect more photons, which is often undesirable under object and/or camera motion. 
Furthermore, CMOS sensors have limited dynamic range. One recent advancement in image sensing technology is the Single-Photon Avalanche Diodes (SPAD) array that is capable of detecting and capturing individual photons~\cite{8449092}, enabling super-fast image readouts on the order of 100,000 frames per second. 
The ability of SPAD sensors to saturate at high intensities makes them suitable for high-dynamic range scenes, and their sensitivity to photons makes them suitable for low-light imaging~\cite{Liu_2022_WACV}.
These advantages provide a compelling reason to swap the traditional RGB color images used by 3D reconstruction techniques with SPAD image captures. 

However, 3D reconstruction from SPAD images remains difficult. 
SPAD images are inherently binary, implying that they provide a poor representation of the visual appearance of the scene. 
Specifically, binary images are heavily influenced by photon noise~\cite{s16071122}, and recovering the scene geometry from such noisy images is challenging. 

Previous work~\cite{jungerman2024radiance} shows the feasibility of NeRF for 3D reconstruction from multi-view SPAD images. However, the synthesized views are heavily smoothened due to averaging effects. Further, the visual appearance is not encoded into the scene, deeming the rendered images ill-suited for several downstream tasks like depth estimation, semantic understanding, etc. Recent methods like Gaussian Splatting or 3DGS capture high-frequency details due to their explicit internal representation and further enable efficient optimization and high-speed rendering at inference time. We argue that 3DGS is perhaps a more well-suited scene representation for reconstruction from SPAD images and can help realize a fast end-to-end system. Although Jungerman and Gupta \cite{jungerman2024radiance} extend their approach to support the GS framework, they do not directly utilize binary images but instead rely on averaging multiple frames. This approach is limited, as the optimal number of frames to average depends on camera motion and requires fine-tuning for each scene, making optimization less efficient.  

% Towards this end, we propose a framework \textit{PhotonSplat}, a novel 3D reconstruction technique from multi-view SPAD image captures, borrowing the internal scene representation from 3DGS. Using the SPAD image formation process, we adapt the rendering process of each Gaussian splat to model the photon detection probabilities at each pixel. This enables our framework to be directly supervised using the binary SPAD images. However, we still observe artifacts in the recovered scene geometry due to the presence of significant noise in the input SPAD captures. Therefore, we add a spatial smoothening that yields higher-quality reconstruction. 

In this work, we present \textit{PhotonSplat}, a novel framework designed to reconstruct scenes captured by a single-photon camera operating at high speeds. In high-speed scene capture, obtaining a high-quality grayscale image through averaging consecutive frames in a sequence presents a significant challenge due to the inherent noise vs. blur tradeoff. This process is scene-dependent, as insufficient averaging results in excessive noise, while over-averaging introduces motion blur. To circumvent this issue, we propose a novel approach that operates directly on binary images, making our approach more robust. By integrating the photon detection process into the splatting framework, we can predict photon probabilities at each pixel, thus enabling explicit scene modeling. To mitigate noise in the binary images, we propose a \textit{3D spatial smoothing filter}, which produces view-consistent, noise-free images. 

The next challenge is to encode the visual appearance of the scene. Directly applying pre-trained colorization models~\cite{kang2023ddcolor} yields inconsistent results that are not desirable. Some existing work proposes hardware-based solutions~\cite{ma2023seeing,gnanasambandam2019megapixel} or requires multiple exposure images~\cite{purohit2024generative}, both of which are not practical for a high-speed capture setup like ours. We propose a view-consistent colorization module that uses a single reference color image that could be significantly motion-blurred to encode the color information of each splat. In cases where obtaining a reference image could be challenging, our framework can easily extend to the use pre-trained 2D priors~\cite{kang2023ddcolor} to encode the scene color. This allows us to render 3D consistent colored images from any arbitrary viewpoint that can be further used for several downstream tasks, including segmentation, instruction-guided scene editing, etc. Additionally, we demonstrate that our method can be adapted to render dynamic scenes with object motion, enabling 4D scene reconstruction. To address the lack of publicly available multi-view single-photon datasets, we introduce \textit{PhotonScenes}, a dataset comprising multi-view captures of nine real-world scenes. We summarize our contributions as follows: 

\begin{itemize}[leftmargin=*]
    \item We propose a gaussian splatting based framework that can recover the underlying scene geometry from multi-view SPAD image captures. 
    \item Introduce a novel 3D Spatial Smoothening filter to mitigate noise in the output renderings. 
    \item The proposed colorization module encodes visual appearances either using a single blurry RGB color image or using pretrained 2D image priors, enabling further downstream tasks like instruction guided editing, and segmentation.
    \item A comprehensive benchmark dataset for evaluating 3D reconstruction on real-world SPAD-captured images, named \textit{PhotonScenes}.
\end{itemize}

% \begin{itemize}
%     \item Novel approach that reconstructs a 3D scene from SPC camera images by modeling photons in 3D to capture scene geometry.
%     \item 
%     \item Introduced \textit{PhotonScenes}, a benchmark of real-world scenes captured using SPC sensors. 
% \end{itemize}
SPAD sensors enable high-speed 3D reconstruction in two primary scenarios: static scenes with rapid camera motion and dynamic scenes with slower motion. In the former, they are particularly valuable for applications such as autonomous vehicles, aerial drones, and robotics, where accurate perception during fast navigation is essential. In the latter, SPADs are well-suited for AR/VR and defense applications, including sports tracking and missile guidance, where capturing fast-moving objects enhances performance and immersion. They also offer advantages in industrial surveillance by detecting transient events that standard CMOS sensors may miss. Our work demonstrates the effectiveness of SPAD-based 3D reconstruction in both static and dynamic settings (see Fig.~\ref{fig:dyn}). As camera technologies evolve, showcasing the utility of emerging sensors like SPADs on tasks traditionally reliant on RGB imagery becomes increasingly important. Despite their limited per-frame information compared to RGB images, we show that SPAD data can robustly recover scene geometry and appearance, offering a promising alternative in challenging conditions such as low light, high speed, and HDR environments.

\section{Related Work}
\label{sec:related_works}

\subsection{Single Photon Imaging}
Single Photon Avalanche Diode (SPAD) camera array is a new sensing technology capable of counting individual photons with precise timing, originally used in active imaging applications like LiDAR~\cite{Kirmani2014-ey},~\cite{Shin2016-du} and fluorescence microscopy~\cite{8449092}. 
Recent improvements, including an increase in spatial resolution of these sensors, make them viable for passive imaging and, along with increased affordability, make them a more practical piece of camera hardware viable for consumer photography.
SPAD images prove to be useful in high-dynamic range scenes~\cite{ingle2019high,liu2022single}, low-light scenarios\cite{goyal2021photon}, non line of sight imaging ~\cite {Buttafava:15},~\cite{OToole2018-bk},~\cite{Callenberg2021CheapSPAD} and high-speed motion~\cite{jungerman2023panoramas,ma2020quanta}. 
% Previous studies have explored the integration of SPAD sensors into RGB SLAM systems~\cite{tofslam} to enhance scene understanding. Some approaches employ supervised learning to extract geometric information from single SPAD measurements, such as reconstructing 3D human poses~\cite{Ruget2022pixels2pose} or generating high-resolution depth maps~\cite{Jungerman_ECCV_22,deltar}. 
In this work, we attempt to harness these benefits for 3D reconstruction by learning to optimize a scene representation from multi-view SPAD images.

\subsection{Novel View Synthesis}
Neural Radiance Fields (NeRFs)~\cite{mildenhall2020nerf} introduced an approach to novel view synthesis and 3D reconstruction, modeling scene characteristics through the weights of a multilayer perceptron (MLP) from posed multi-view images. Since then, significant enhancements have addressed NeRF’s limitations under challenging imaging conditions. For instance, methods in \cite{ma2022deblurnerf,wang2023badnerf,peng2023pdrf} focus on generating sharp novel view renderings from blurry RGB inputs, while techniques in \cite{cui2024aleth,wang2023lighting} specialize in low-light image conditions. Despite these advancements, NeRFs are slow to train and render as they need to query the MLP for each point sampled on the ray to calculate the point-wise density and color.

Recently, Gaussian Splatting (GS) has gained traction for its rapid training and rendering capabilities, utilizing a CUDA-optimized rasterization pipeline that supports real-time rendering. This framework has seen considerable expansion; works such as \cite{wu20244d,duan20244d} extend GS to incorporate dynamic scenes, while \cite{guedon2024sugar,huang20242d} adapt the GS pipeline for high-quality mesh extraction. Additionally, \cite{chen2024deblur,lee2024deblurring} introduce camera trajectory estimation to mitigate blur in rendered images from motion-corrupted inputs. These advancements, alongside other modality-based approaches, continue to improve robustness and quality in real-world 3D scene synthesis and novel view applications.

\subsection{Novel View Synthesis for other sensor modalities}
Recent efforts have expanded Neural Radiance Fields (NeRFs) to encompass diverse imaging modalities, including thermal, hyperspectral, event-based, lensless, and single-photon data. For example, thermal scene reconstruction has been achieved by integrating thermal image datasets with NeRF-based models~\cite{ye2024thermal,lin2024thermalnerf,lu2024thermalgaussian}. 
% In hyperspectral imaging, \cite{chen2024hyperspectral} presented a method to produce a hyperspectral 3D reconstruction where each spatial point and viewing angle includes wavelength-dependent radiance and transmittance data. 
Event-based data has shown promise in handling high-speed motion, with methods like Event3DGS~\cite{xiong2024event3dgs} and E2GS~\cite{deguchi2024e2gs} achieving high-fidelity 3D structure and appearance reconstruction under rapid ego-motion. Lensless imaging approaches, such as the method developed by \cite{madavan2024ganesh}, enable clean novel-view rendering from multi-view lensless image captures. 
Moreover, advancements in single-photon imaging have been demonstrated by Quanta Radiance Fields (QRF)\cite{jungerman2024radiance}, which proposed a framework that takes binary images from single-photon cameras to generate novel-view renderings even in high-speed scenarios. However, their approach faces limitations in training and rendering speed, primarily due to the computational intensity inherent to NeRFs. Our proposed method addresses these efficiency challenges by achieving faster training and rendering speeds, advancing the practical use of single-photon camera data for novel-view synthesis. 

\subsection{Gaussian Splatting}
\label{sec:preliminary}

Our framework builds upon GS~\cite{kerbl3Dgaussians}, which represents a 3D scene using a collection of anisotropic 3D Gaussians.
Below, we equip the reader with the necessary background. 
Given $N$ multi-view images, their corresponding camera poses $\{{I}_{i}, {P}_{i}\}_{i=1}^{N}$, and a sparse point cloud, we initialize a Gaussian point at a given location ${x}$ as
\begin{linenomath*}
\begin{equation}
\label{formula:gaussian_formula}
    \mathcal{G}({x}) = e^{-\frac{1}{2}{(x-\mu)}^T{\Sigma}^{-1}{(x-\mu)}},
\end{equation}
\end{linenomath*}
where ${\Sigma}$ is the covariance matrix and ${\mu}$ is the center position of Gaussian. 
To render an image, each Gaussian is first projected onto the image plane using the camera intrinsic, followed by alpha blending to synthesize the final color image. 
Mathematically, the color ${C}$ of each queries pixel ${x'}$ is given by:
\begin{linenomath*}
\begin{equation}
\label{formula:splatting_volume_rendering}
    {C}({x'}) = \sum_{k \in M} {c}_{k} {\alpha}_{k} \prod_{j=1}^{k-1} (1-{\alpha}_{j}),
\end{equation}
\end{linenomath*}
where $M$ denotes the number of Gaussians corresponding to this pixel, and ${c}_{k}$, ${\alpha}_{k}$ are other Gaussian attributes representing the view-dependent color and opacity of each point. 
The individual attributes are optimized to match the rendered color with the target image using the $\mathcal{L}_{\text{1}}$ and $\mathcal{L}_{\text{SSIM}}$, i.e., the L1 loss and D-SSIM losses.
\begin{linenomath*}
\begin{equation}
\label{formula:loss_preliminary}
    \mathcal{L} = (1 - \gamma)\mathcal{L}_{\text{L1}}(\widehat{{C}}_{\text{target}}, {C}_{\text{target}}) + \gamma \mathcal{L}_{\text{SSIM}}(\widehat{{C}}_{\text{target}}, {C}_{\text{target}}),
\end{equation}
\end{linenomath*}
where $\gamma$ is a scaling factor, $\widehat{{C}}_{\text{target}}$, and ${C}_{\text{target}}$ denote the final rendered image and ground truth image from the target viewpoint respectively. 

% The rendering process of Gaussian splatting involves two primary stages. Initially, each 3D Gaussian is projected onto the 2D image plane, resulting in a 2D Gaussian representation. This transformation is accomplished by projecting the 3D Gaussian's mean to the 2D space, while the covariance matrix is calculated using a viewing transformation combined with the Jacobian matrix from an affine approximation to approximate the 2D covariance.

% For the color rendering of each pixel, alpha blending is applied to a sequence of $N$ 3D Gaussians ordered from front to back. The pixel color $C$ is then computed as:
% \begin{equation}
% \label{formula:splatting_volume_rendering}
%     C = \sum_{i \in N} c_i \alpha_i \prod_{j=1}^{i-1} (1-\alpha_j),
% \end{equation}
% where each $\alpha$ is derived by modulating the opacity $o$ with the influence of the 2D covariance contributions from $\Sigma$ and the projected image plane coordinates. The color $c_i$ is represented using Spherical Harmonics (SH) to capture view-dependent effects, while the covariance matrix $\Sigma$ is characterized by a unit quaternion $q$ for rotation and a scaling vector $s$ in 3D. This setup enables efficient, realistic scene representation and facilitates various downstream tasks.

\section{Method}
\label{sec:method}

\begin{figure*}[h!]
  \includegraphics[width=\textwidth]{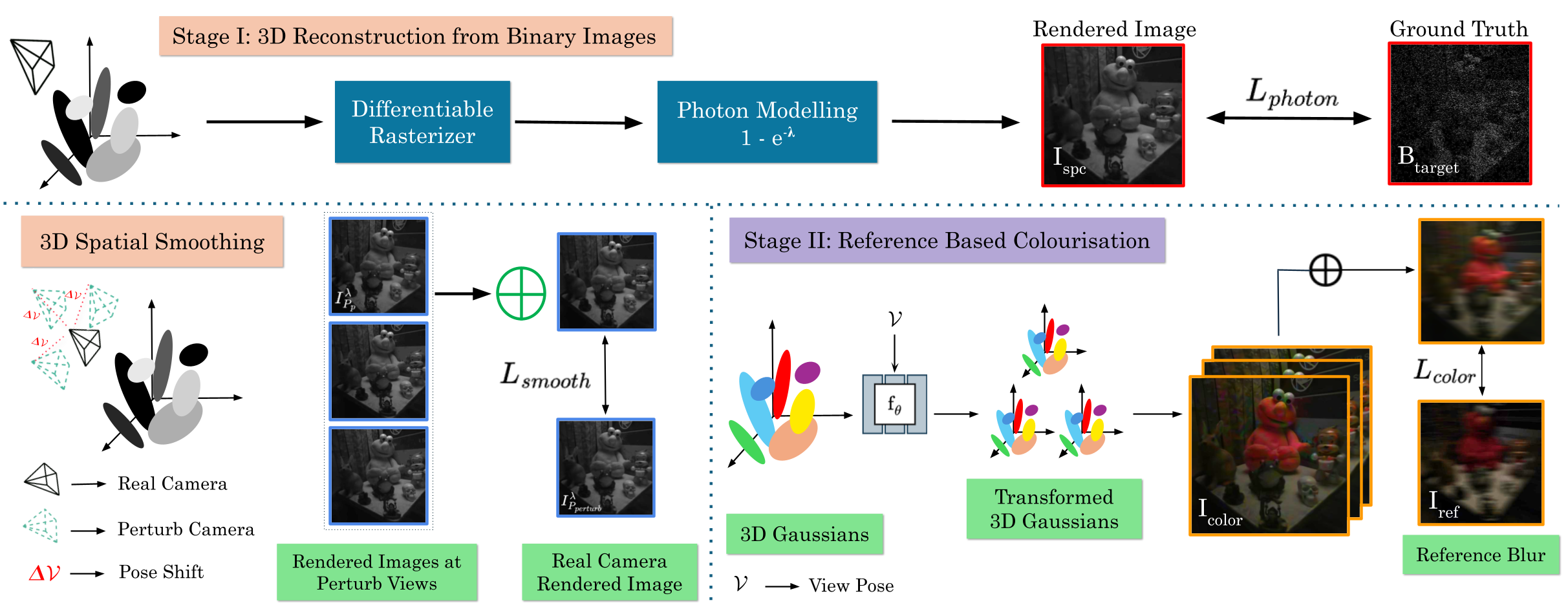}
\caption{\textbf{Method Overview. } PhotonSplat learns to recover a 3D scene from binary SPAD images. We incorporate the photon hitting probabilities directly into the Gaussian splat enabling it to model the SPAD image formation process. In addition to a smoothening regularization(part of Stage I \& Stage II), we can accurately recover the scene geometry. Finally, our colorization module jointly models camera motion and color attributes, enabling it to encode the visual appearance of a scene with a single reference blurry image. 
% Our objective is to reconstruct 3D scenes and render novel views from binary images. The GS model estimates photon-hitting probabilities by rendering $\lambda = \phi * \tau$ and applying the photon modeling equation. We optimize this using a binary cross-entropy loss between the rendered probabilities and the binary target image. To reduce overfitting to photon noise, we introduce a perturbation loss, comparing the rendered image with averaged outputs from perturbed camera views, ensuring smoother results. Additionally, a network learns transformations for the 3D Gaussians, rendering and averaging transformed views, which are supervised using ground truth blurry images to jointly deblur and colorize the scene.
 }
\label{fig:modelarch}
\end{figure*}

\begin{figure}[h!]
  \includegraphics[width=0.48\textwidth]{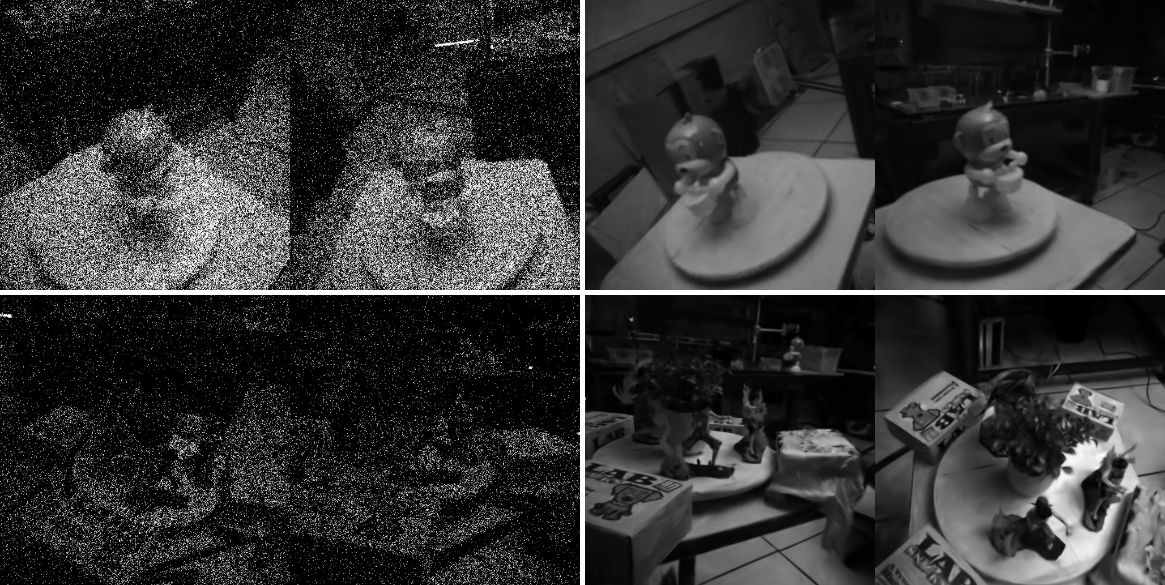}
  \caption{The first 2 columns represent our dataset that includes multi-view SPAD images captured under diverse lighting and camera motions ensuring a robust evaluation framework. Last 2 columns represent novel-view renderings from our methodology. }
  % The top row represents well-lit scenes, while the bottom row exhibits low-light conditions, evident from sparse white pixels. Left-column scenes involve aggressive camera motion, whereas right-column scenes follow a forward-facing capture. This diversity ensures a robust evaluation framework 
  \label{fig:dataset}
\end{figure}

%\textbf{Overview.} 
We introduce \textit{PhotonSplat}, a framework that reconstructs a 3D scene from its corresponding multi-view SPAD image captures. 
Formally, given $N$ multi-view calibrated SPAD image captures with their corresponding pose information $\{{B}_{i}, {P}_{i}\}_{i=1}^{N}$, we aim to learn the Gaussian attributes that recover both scene geometry and visual color. 
Using the optimized Gaussian scene, we can synthesize clean, colored novel views from any arbitrary angle that can prove useful for several downstream tasks like instruction-guided editing, segmentation, etc. 
We first adapt the Gaussian attribute to now represent photon detection probabilities using the image formation model of SPAD sensors (Sec. \ref{sec:modelphotons}). SPAD images are heavily distorted by photon noise, resulting in view inconsistency and introducing significant noise into the optimized Gaussians.
We therefore introduce a spatial smoothening regularization that alleviates this problem (Sec. \ref{sec:spatialfilter}). 
SPAD images contain no information regarding the color appearance of the scene. 
Assuming access to either a single blurred color image or generative priors~\cite{kang2023ddcolor}, we can now encode the visual appearance of the scene using a colorization module introduced in Sec. \ref{sec:3dcolor} and outline the optimization strategy in Sec.\  \ref{sec:training}. 
Our overall pipeline is indicated in Fig. \ref{fig:modelarch}. 

% \textbf{Overview.} In this work, we introduce \textbf{PhotonSplat}, a framework for reconstructing 3D scenes from binary images captured by high-speed single-photon cameras. Given a set of $N$ multi-view binary images and camera poses obtained via Structure-from-Motion (SfM), our aim is to generate high-quality, denoised novel views from arbitrary perspectives and colorize the scene. Building on the Gaussian Splatting (GS) method (Sec. \ref{sec:preliminary}), we adapt it to predict pixel-wise photon impact probabilities instead of traditional scene radiance, enabling effective 3D reconstruction directly from binary images avoiding noise-vs-blur trade-off (Sec. \ref{sec:modelphotons}). To address the noise in binary captures, we propose a 3D spatial smoothing filter (Sec. \ref{sec:spatialfilter}) that significantly improves model quality by reducing noise. Additionally, we extend the framework to enable scene colorization through two strategies: reference-based     colorization guided by a single blurry RGB image (Sec. \ref{sec:refcolor}) and no-reference colorization using 2D priors (Sec. \ref{sec:norefcolor}). We outline and the final training strategy and the optimisation equation in Sec. \ref{sec:finaloptimisation}.

\subsection{Modeling Photons in 3D}
\label{sec:modelphotons}

When a photon is incident on a SPAD image pixel, it starts an avalanche reaction that triggers a detection event. This is a form of photon counting which avoids any form of read noise unlike traditional camera hardware. Given a scene with radiant flux ${\phi}$, the number of incident photons $n$ arriving in time ${\tau}$ follows a Poisson distribution~\cite{jungerman2024radiance} given by
\begin{linenomath*}
\begin{equation}
    P(n) = \frac{({\phi} {\tau})^n e^{-{\phi} {\tau}}}{n!}.
\end{equation}
\end{linenomath*}
SPAD pixels are designed to count the occurrence of a single photon, resulting in binary observations. 
Simply put, a pixel measurement ${b}$ is given the value one if one or more photons are incident in time ${\tau}$ and zero otherwise.
This follows a Bernoulli distribution given by
\begin{linenomath*}
\begin{equation}
    \begin{aligned}
        P({b}=0) &= P(n=0) = e^{-{\lambda}} \,, \text{with ${\lambda} = {\phi} {\tau}$}\\
        P({b}=1) &= P(n \ge 1) = 1 - e^{-{\lambda}}.
    \end{aligned}
\end{equation}
\end{linenomath*}
Hence, images captured by a SPAD sensor are fundamentally affected by photon noise, making them ill-suited for 3D reconstruction. In our work, the Gaussian splats are optimized to estimate ${\lambda}$, which is not binary in nature and simpler to learn. This enables us to estimate the radiance field better and model the photon hitting probabilities in a more physics-informed manner.
More specifically, we modify Eq. \ref{formula:splatting_volume_rendering} and estimate ${C}_{gray}$ at each projected 2D location as:
\begin{linenomath*}
\begin{equation}
\label{formula:gray}
    {C}_{gray}({x'}) = \sum_{k \in M} {{c}_{gray}}_{k} {\alpha}_{k} \prod_{j=1}^{k-1} (1-{\alpha}_{j}),
\end{equation}
\end{linenomath*}
where ${{c}_{gray}}_{k}$ is the Gaussian attribute representing its contribution towards ${\lambda}$ at each point. 
Subsequently, we can now supervise the rendered binary frames ${I}^{\text{SPAD}}_{\text{target}}$ with the target binary frame ${B}_{\text{target}}$ using the binary cross entropy loss as:
\begin{linenomath*}
\begin{equation}
    \mathcal{L}_{\text{photon}} = \mathcal{L}_{\text{BCE}}({I}^{\text{SPAD}}_{\text{target}}, {B}_{\text{target}}) \text{, where } {I}^{\text{SPAD}}_{\text{target}} = 1 - e^{-{C}_{gray}}.
\end{equation}
\end{linenomath*}
This, in turn, provides supervision to all Gaussian attributes, including ${\alpha}_{j}$, Gaussian position ${x}$, which determine the underlying scene geometry. Instead of directly predicting the color value $C \in [0, 1]$ as in the original GS formulation, we modify the model to predict $\lambda$, represented by $C_{\text{gray}} \in [0, \infty)$ in our formulation. This formulation incorporates physical principles into GS, enabling accurate photon modeling and high-quality results.

\begin{figure*}[h!]
\centering
  \includegraphics[width=0.95\textwidth]{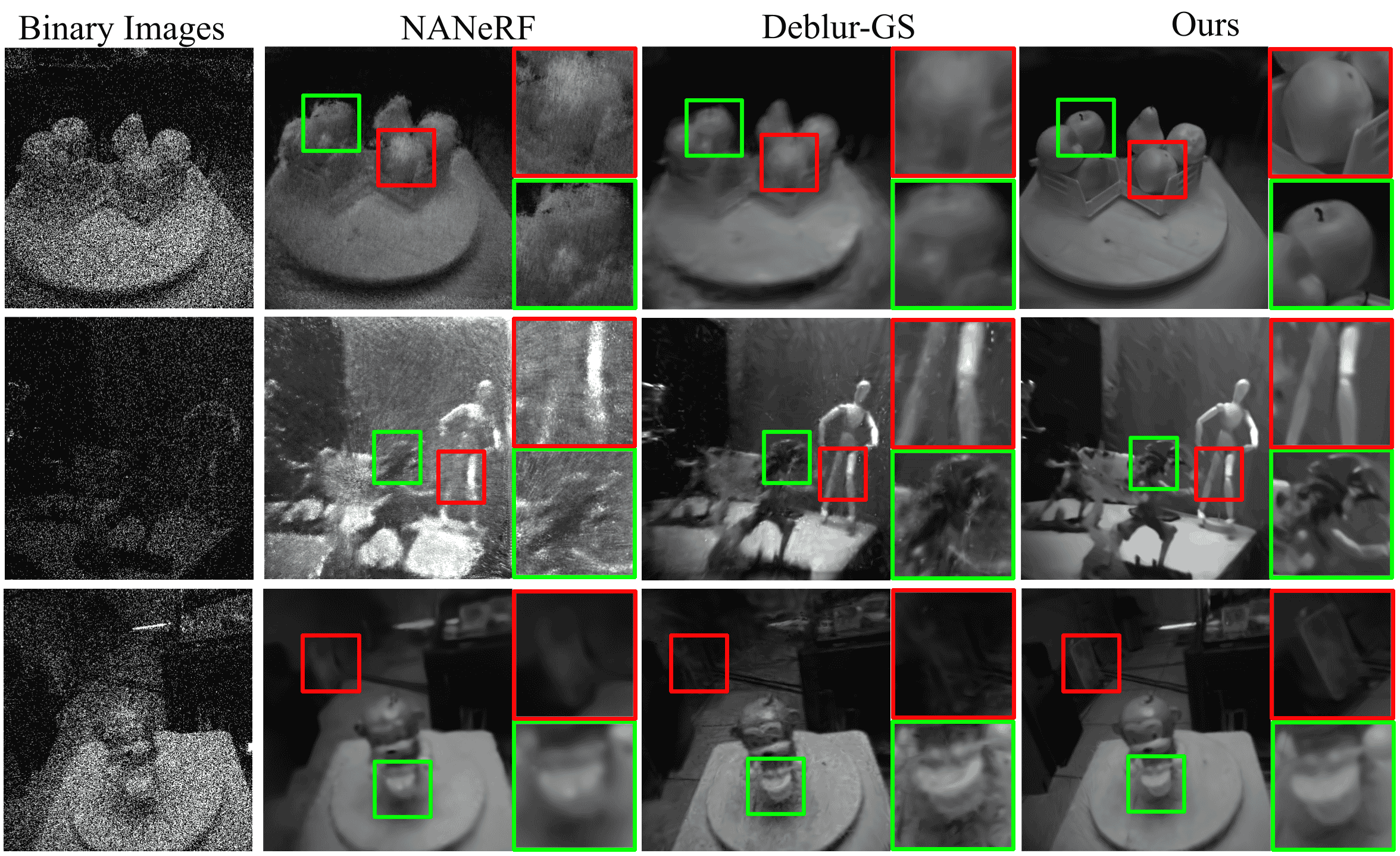}
\caption{Qualitative results for novel view synthesis from multi-view binary frames on our real-world \textit{PhotonScenes} dataset. Baselines such as NANeRF and Deblur-GS fail to reconstruct scenes with high geometric fidelity, either oversmoothing details or retaining noise. In contrast, PhotonSplat accurately reconstructs individual fruits (row 1) and captures fine details of toys (rows 2 and 3).}
  \label{fig:recon}
\end{figure*}

\vspace{-3mm}
\subsection{Spatial Smoothing Regularization}
\label{sec:spatialfilter}
The raw images from SPAD arrays are quantized and heavily influenced by photon noise. This leads to severe artifacts in the learned geometry (see Fig. \ref{fig:abl}). Gaussian Splatting has a tendency to overfit to training views, which further increases the noise present in novel view renders.
We introduce a simple yet effective regularization to ensure that nearby views generate smooth outputs, implying that the noisy components will be automatically removed. 
Given a randomly selected pose ${P}_{\text{perturb}}$, we perturb the translation matrix with a small amount of noise to obtain nearby viewpoints. 
Next, we minimize the difference between the rendered ${I}^{\text{SPAD}}$ images of the selected pose and its nearby perturbed views. 
Formally, this is given by:
\begin{linenomath*}
\begin{equation}
\label{formula:smooth}
    \begin{aligned}
        \mathcal{L}_{\text{smooth}} = \mathcal{L}_{\text{L1}} \left( \frac{1}{p} \sum_{p} {I}^{\text{SPAD}}_{{P}_{p}}, {I}^{\text{SPAD}}_{{P}_{\text{perturb}}} \right), \\
        \text{where } p \in \{{P}_{\text{perturb}} + \mathcal{N}(0, \sigma)\}.
    \end{aligned}
\end{equation}
\end{linenomath*}
Here $\mathcal{N}(0, \sigma)$ indicates a Normal distribution with standard deviation $\sigma$ indicating the strength of smoothness.
A higher value generates more smooth renderings compared to a smaller value, and we obtain an optimal value through several experiments. The proposed approach eliminates the need for scene-specific fine-tuning by employing a per-frame regularization strategy that operates independently of camera motion. Notably, a fixed $\sigma$ parameter demonstrates strong generalization capability across a wide range of scenes (see Fig. \ref{fig:abl_spatial}).

\vspace{-2mm}
\subsection{View-Consistent Colorization}
\label{sec:3dcolor}

Given that we have successfully recovered the underlying geometry, the next important step is to encode the color appearance of the scene. However, the input SPAD images do not contain any information regarding the color. It is, however, easy to capture a single blurred color image captured using conventional RGB cameras moving at the same speed as the original SPAD sensor. 
This could provide useful information to learn the color attributes of each Gaussian splat. 

Given a reference color image ${I}_{\text{ref}}$ and its corresponding pose ${P}_{\text{ref}}$, naively supervising the rendered views could yield suboptimal results since the reference image is in itself severely motion-blurred. 
Instead, we propose to jointly learn the camera motion along with the color attributes, which together can be used to render a blurred image at the reference viewpoint. 
Such disentanglement ensures that accurate color can be propagated into individual splats, improving visual quality.

Inspired by previous works on spline-based deblurring\cite{lee2024deblurring, chen2024deblur}, formally we estimate $m$ deformations for each Gaussian attribute using a neural network. Next, we re-render the image using each individual deformation from the reference viewpoint and average the synthesized images to simulate motion blur due to camera movement. 
We modify Eq. \ref{formula:loss_preliminary} as follows:
\begin{linenomath*}
\begin{equation}
\label{formula:color}
\begin{aligned}
    \mathcal{L}_{\text{color}}({a}, {b}) = (1 - \gamma)\mathcal{L}_{\text{L1}}({a}, {b}) + \gamma \mathcal{L}_{\text{SSIM}}({a}, {b}), \\
    \text{where } a = \frac{1}{m}\sum_{l=1}^{m}{C}_{color} \text{ and } b = {I}_{\text{ref}}
\end{aligned}
\end{equation}
\end{linenomath*}
where ${C}_{color}$ indicates the rendered color image after individual deformations.

We observe that ${C}_{gray}$ contains continuous values unlike the original binary SPAD images and resemble images close to a grayscale image. 
This allows us to couple the SH(spherical harmonics) parameters of ${C}_{gray}$ and ${C}_{color}$ using the standard linear transformation from color to grayscale images. Moreover, this relationship enables superior quality in the learned visual appearances even with limited supervision, i.e., a single reference image. Note that Eq.~\ref{formula:gray} reconstructs the scene’s geometry in grayscale without visual appearance, while Eq.~\ref{formula:color} enables the addition of color to the scene.

\noindent \textbf{Without a reference image. }In several instances, it might be difficult to procure a reference image suitable for encoding such visual appearances. 
However, we can leverage existing pre-trained 2D generative priors to hallucinate color attributes for each splat. 
Specifically, we apply a pre-trained DDColor~\cite{kang2023ddcolor} model to estimate a colored reference view and proceed with the steps discussed above. 
However, in practice, we observe that DDColor cannot successfully colorize the renders from the first stage due to the presence of minor noise in the synthesized views. 
Therefore, we apply a post-processing step that first denoises the rendered reference view using a UNet-based architecture~\cite{wang2018high}, followed by colorization, before optimizing the splat attributes using the same. 
Our colorization module can, therefore, successfully encode visual appearances with or without a reference color view - thereby improving the practical applications of our proposed framework. Refer to supplementary for more details.

\subsection{Training and Inference}
\label{sec:training}
In each training iteration, we optimize for scene geometry and visual appearance in two stages. 
The first stage learns to reconstruct the underlying scene geometry, and the overall loss criterion is as follows:
\begin{linenomath*}
\begin{equation}
\label{formula:loss_stage1}
    \mathcal{L}_{\text{stage1}} = \mathcal{L}_{\text{photon}} + \mathcal{L}_{\text{smooth}}.
\end{equation}
\end{linenomath*}
The second stage optimizes for $\mathcal{L}_{\text{color}}$ to bake visual appearances into the scene representation.
We represent these individual stages diagrammatically in Fig. \ref{fig:modelarch}. Here, $\mathcal{L}_{\text{photon}}$ facilitates geometric reconstruction, while $\mathcal{L}_{\text{color}}$ incorporates texture details into the scene. This process is performed sequentially, as simultaneous optimization of both geometry and color leads to model instability. This instability may arise from optimizing gaussians using both the binary cross-entropy loss and the L1 loss, potentially increasing the number of gaussians to a very high value. To ensure stable training, we couple the spherical harmonic (SH) parameters of grayscale and color Gaussians through a color-grayscale transformation.

\vspace{-3mm}
\section{Experiments}

% \begin{figure*}[h!]
% \centering
%   \includegraphics[width=0.95\textwidth]{figures/main_3drecon_image_cropped.png}
% \caption{Qualitative results for novel view synthesis from multi-view binary frames on our real-world \textit{PhotonScenes} dataset. Baselines such as NANeRF and Deblur-GS fail to reconstruct scenes with high geometric fidelity, either oversmoothing details or retaining noise. In contrast, PhotonSplat accurately reconstructs individual fruits (row 1) and captures fine details of toys (rows 2 and 3).}
%   \label{fig:recon}
% \end{figure*}

\begin{figure*}[h!]
\centering
  \includegraphics[width=0.9\textwidth]{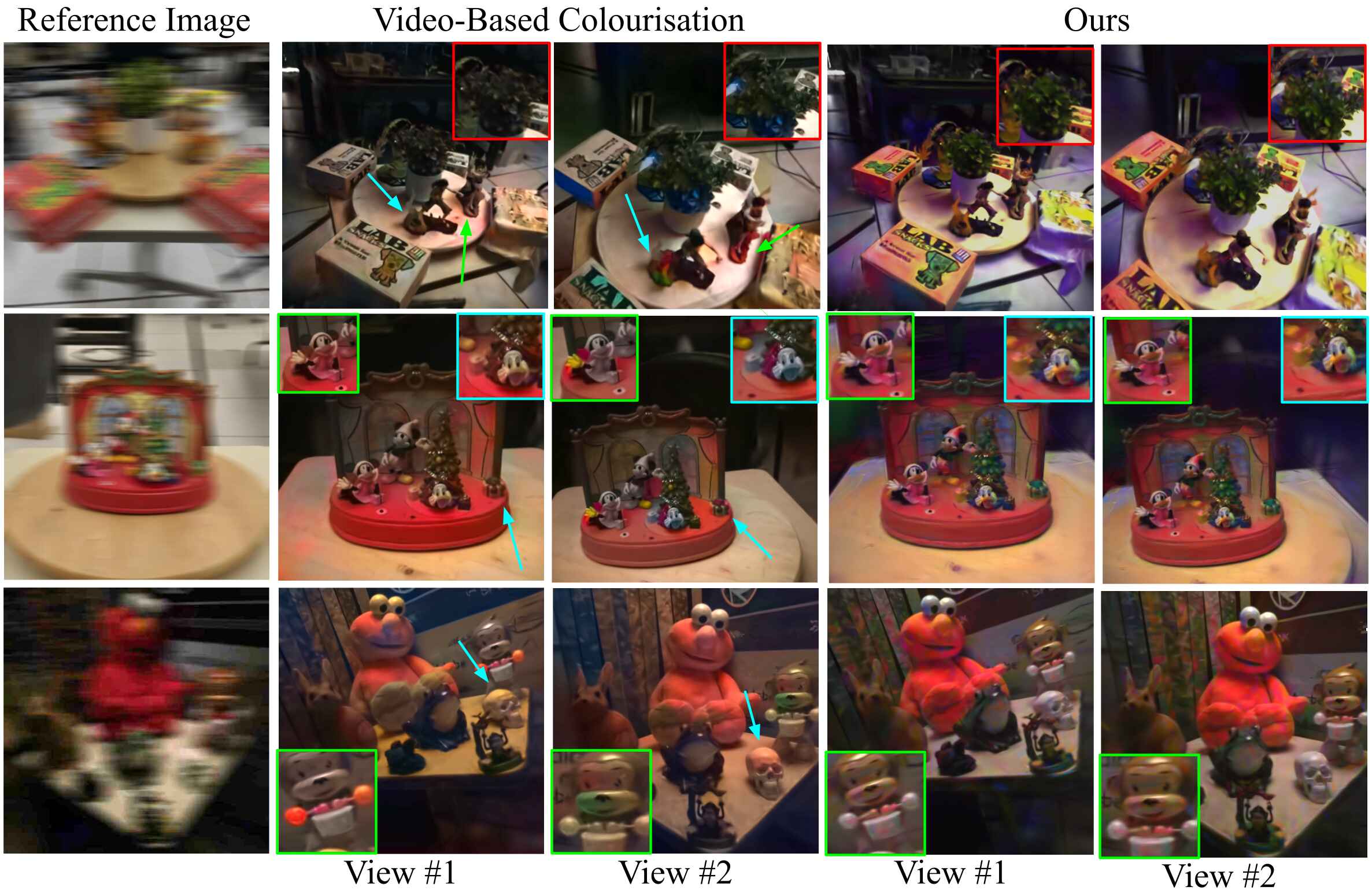}
\caption{Qualitative results for colorization based on reference image on the \textit{PhotonScenes} Dataset. PhotonSplat achieves consistent scene colorization (row 1 and row 3) and closely resembles the reference image (row 2) better than baselines. }
  \label{fig:refcolor}
  % \vspace{-3em}
\end{figure*}

We conduct several experiments to evaluate the efficacy of \textit{PhotonSplat} to reconstruct a 3D scene from multi-view SPAD captures. 
We showcase results on both simulated and real SPAD captures and discuss them in more detail below. 

\vspace{-3mm}
\subsection{Implementation Details}
\label{sec:implementation}
Our implementation builds upon the original 3D Gaussian Splatting  framework~\cite{kerbl3Dgaussians}. We incorporate the additional loss functions described in Sec. \ref{sec:method} and train each scene representation for about 20,000 iterations. All hyperparameters—loss weights, camera perturbation variance for spatial smoothing, and training iterations—were tuned on a single scene. We found that a range of $\sigma$ values perform well, indicating the method is not sensitive to the value of $\sigma$. $\sigma$ in $\mathcal{L}_{\text{smooth}}$ is set to 0.0005, $\gamma$ in $\mathcal{L}_{\text{color}}$ is set to 0.2 and we render 3 nearby viewpoints. Additionally, we observe that the smoothening regularization is best applied after few training steps, which we set to 15,000 iterations. In all our experiments unless otherwise stated, we use a single blurry reference image as prior to encode visual appearances, and set $m$ to 4 to learn deformations that capture the camera motion. 

In our experiments across 9 scenes, SfM successfully registered camera poses in 8. It struggled in one scene (see Fig. 10) captured under extremely low-light conditions. Although camera motion varied across scenes, we believe the low-light environment was the main factor affecting performance, posing a greater challenge than motion blur. All the models are trained on an RTX 4090 GPU, requiring about 5-10 mins to optimize for each scene. We employ Uformer, a U-Net-based architecture, for denoising, leveraging its hierarchical encoder-decoder structure with LeWin Transformer blocks. For training details, including hyperparameters, refer to the supplementary material. To align binary and color frames, we grayscale and center-crop the color image to match the binary resolution, then jointly register poses using SfM. Color frames were captured separately using a CMOS camera moving at a similar speed as the SPAD, without stereo or beam-splitter setup. We also recorded sharp ground-truth images at matching positions, as shown in Fig. E of the supplementary alongside the blurry frames.
% incorporating a binary cross-entropy loss to optimize scene reconstruction from binary images. Camera poses are obtained through Structure-from-Motion (SfM) techniques, specifically COLMAP \cite{fu2023colmap} and GLOMAP \cite{pan2024glomap}. Since these methods are not designed for binary images, we pre-process the data by averaging frames and removing the dead pixels using median filtering, allowing SfM to generate the initial point cloud and camera poses. While averaging can introduce motion blur, this does not significantly affect pose estimation, as minor blur has a negligible impact on feature matching. For 3D spatial filtering, we introduce perturbation by varying the translation component of the camera matrices with a variance of 0.0005, simulating 3 additional viewpoints. We train the entire model for 20000 iterations and start the spatial filter loss only after 15000 iterations. For the colorization task, we utilize 2-3 reference blurry RGB images, which are shown to the model approximately 100 times per epoch, ensuring sufficient exposure for the model to learn colorization effectively. All models are trained on an NVIDIA RTX 4090 GPU, with each scene requiring approximately 5-10 minutes for training.

% \begin{figure}[h!]
%   \includegraphics[width=0.45\textwidth]{figures/setup.jpeg}
%   \caption{\textbf{Single Photon Camera Setup.}}
%   \label{fig:setup-main}
% \end{figure}

\vspace{-1mm}
\subsection{Datasets}
\label{sec:data}

We evaluate our method on both simulated and real SPAD captures. We discuss them below.

\textbf{Simulated Captures. } From existing multi-view RGB datasets, we generate synthetic scenes that simulate SPAD images\cite{mildenhall2020nerf}. Following the data generation strategy from \cite{goyal2021photon}, we first simulate photon-starved imaging by scaling pixel intensities, followed by applying a Poisson process to model photon arrival. Finally the photon counts are thresholded to create binary image captures, closely emulating real scans. We capture 15 scenes using a handheld camera set on a higher framerate to more closely replicate the image readout speeds of a SPAD camera.

\begin{figure*}[ttt]
\centering
  \includegraphics[width=0.95\textwidth]{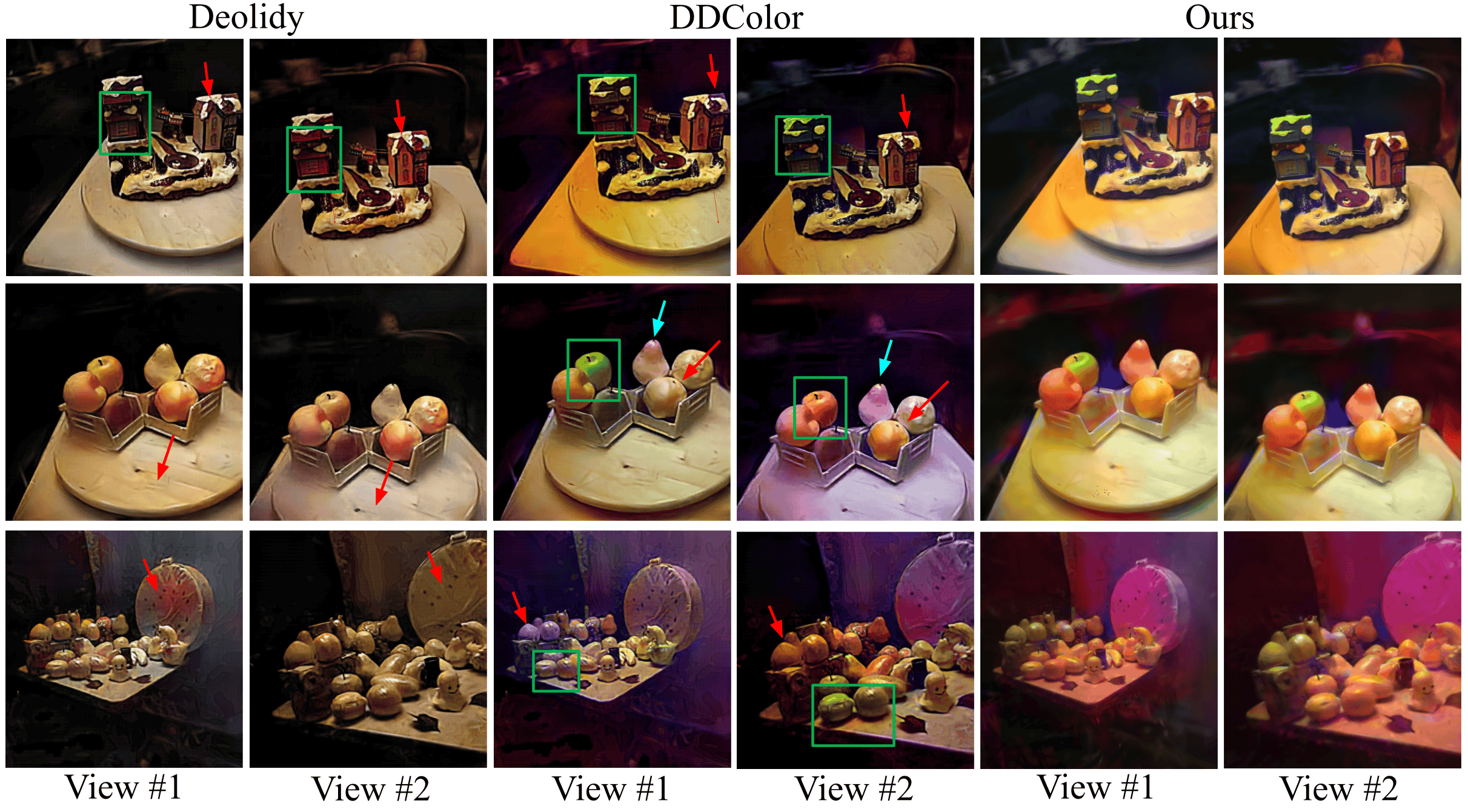}
\caption{Qualitative results for colorization without any reference image on the \textit{PhotonScenes} dataset. Our framework achieves consistent scene colorization, outperforming other comparison methods. We show arrows in the figures to point out the inconsistencies present in the baseline.} 
  \label{fig:norefcolor}
  % \vspace{-3em}
\end{figure*}

\begin{table}[ttt!]
    \centering
    \caption{Reconstruction quality on simulated captures. The \sethlcolor{black!30}\hl{best} scores and \sethlcolor{black!15}\hl{second best} scores are highlighted with their respective colors. }
    \label{tab:3dreconsyn}    \resizebox{0.7\columnwidth}{!}{
    \begin{tabular}{lccc}
        \toprule
        Models & \hspace{0.2em}PSNR$\uparrow$\hspace{0.2em} & SSIM$\uparrow$\hspace{0.2em} & LPIPS$\downarrow$\hspace{0.2em}\\
        \midrule
        NANeRF & 13.46 & 0.346 & 0.662\\ 
        Deblur-GS & \cellcolor{black!15}14.50 & \cellcolor{black!15}0.596 & \cellcolor{black!15}0.493\\ 
        \midrule
        Ours & \cellcolor{black!30}15.61 & \cellcolor{black!30}0.737 & \cellcolor{black!30}0.292\\
        \bottomrule
    \end{tabular}
    }

\end{table}

\begin{table}[ttt!]
    % \vspace{-3em}
  \centering
   \caption{Consistency metrics for colorization on simulated captures}
  \label{tab:consis}
  \resizebox{\columnwidth}{!}{
  \begin{tabular}{lcccc}
    \toprule
    Method & Short Range Consistency$\downarrow$ & Long Range Consistency$\downarrow$\\
    \midrule
    Image & 0.060 / 0.136 & 0.108 / 0.208\\
    Video & \cellcolor{black!15}0.052 / \cellcolor{black!15}0.116 & \cellcolor{black!15}0.083 / \cellcolor{black!15}0.163\\
    \midrule
     Ours & \cellcolor{black!30}0.046 / 0.106 & \cellcolor{black!30}0.071 / 0.146\\
    \bottomrule
  \end{tabular}
  }

    % \vspace{-2em}
\end{table}

\textbf{Real Captures. } For real SPAD captures, we use a SPAD512² sensor from PI Imaging~\cite{spadcamera}, as shown in Fig. 1. 
To avoid saturation, we use dim lighting that also preserves the quality of the binary images better. The SPAD sensor captures 512×512 images at 130,000 fps. Despite the high frame rate, only 3,000–5,000 frames were used for training and evaluation. Data was collected under varied lighting to ensure robustness. On average, 12.2\% of photons were detected per binary frame, with a standard deviation of 8.36.
We capture a total of 9 scenes, each with varying complexity arising from occlusions, textures, etc. 
We call this dataset \textit{PhotonScenes} and showcase few scenes in Fig. \ref{fig:dataset}. 

For each scene, we obtain the camera poses through Structure-from-Motion (SfM) techniques, specifically COLMAP~\cite{fu2023colmap} and GLOMAP \cite{pan2024glomap}. 
Neither of these methods are originally implemented for noisy binary SPAD images. 
By averaging multiple frames and removing dead pixels using median filtering, we obtain a rather coarse grayscale version of each view. 
Although smoothened, when fed into the SfM pipeline, they are sufficient to generate a reasonable estimate of camera poses and a sparse point cloud required to initialize the Gaussian splat. 

% \textbf{PhotonScenes Dataset: } We present PhotonScenes, a real-world dataset captured using the SPAD512² sensor from PI Imaging \cite{spadcamera}, as shown in Fig. \ref{fig:setup-main}. To prevent sensor saturation and preserve the quality of binary images, dim lighting was used, leveraging the camera's high sensitivity. The dataset includes 9 diverse scenes, featuring a variety of colors, occlusions, and intricate textures, with each scene containing approximately 1000-4000 multi-view images. All qualitative results in this paper are evaluated on this dataset, demonstrating the robustness of our model to real-world scenes. 

% \textbf{Synthetic Dataset: } For quantitative analysis, we generate synthetic single-photon captures from multi-view RGB datasets. Following the approach in \cite{goyal2021photon}, we simulate photon-starved imaging by scaling pixel intensities, applying a Poisson process to model photon arrival, and thresholding photon counts to create binary frames. This process emulates the binary outputs of the SPAD sensors. We collected 15 scenes using a slow-motion camera to approximate the high temporal resolution of SPC as closely as possible.
\vspace{-1mm}
\subsection{View Synthesis from SPAD Images}

As discussed before, SPAD images are binary and severely prone to photon noise. 
Averaging multiple frames reduces the noise but yields smoothened images. 
Therefore, the problem of view synthesis from SPAD images can be viewed in two perspectives, either jointly denoise and render novel views directly from the SPAD image captures or jointly deblur and render novel views from averaged SPAD images.
This motivates us to create two baselines to compare against (1) using NANeRF~\cite{pearl2022nan}, a novel view renderer from noisy images, and (2) Deblur-GS~\cite{lee2024deblurring}, designed to deblur and synthesize views from any arbitrary angle. We do not have access to ground truth RGB images and, therefore we simply compare the image quality metrics of the rendered and ground truth views in grayscale to evaluate the recovered geometry. 
Table. \ref{tab:3dreconsyn} qualitatively presents these results for the simulated scenes where we have access to the ground truth RGB images. 
Fig. \ref{fig:recon} presents the same qualitatively for real SPAD captures. 
We can clearly see that our method showcases improvements across all three metrics (from Table. \ref{tab:3dreconsyn}), which is also visually indicated in Fig. \ref{fig:recon}. 
Compared to the baselines NANeRF and DeblurGS, our method can maintain an ideal balance between smoothening noise and capturing high-frequency details. 
This indicates that directly modeling the photon hitting probabilities internally in the scene representation enables better recovery of scene geometry compared to others. 

% Binary images captured by SPAD cameras are inherently noisy, and simple averaging introduces a trade-off between noise and blur. Averaging fewer frames retains noise but avoids blur while averaging more frames reduces noise at the cost of motion blur. To address this, we compare our approach against two baselines: NANerf \cite{pearl2022nan}, a NeRF-based noise removal method, and Deblur-GS \cite{lee2024deblurring}, designed to deblur input frames. We evaluate on the PhotonScenes dataset (Sec. \ref{sec:photonscenes}) and present results in Fig. \ref{fig:nvs_binary}. PhotonSplat effectively reconstructs the scene, mitigating noise and preserving high-quality details. In contrast, both NANerf and Deblur-GS struggle with high-speed captures. NANerf fails to remove the slight motion blur introduced by averaging 4 frames, while for Deblur-GS, we average around 64 frames, but still, the model overfits to residual noise, resulting in artifacts in novel-view renders. PhotonSplat directly processes binary images, leveraging a 3D Spatial Filter to eliminate noise and generate clean, high-quality renders. We also perform quantitative evaluations on synthetic datasets (Table \ref{tab:nvs_binary}), where our method outperforms the baselines in all three metrics. These results validate the efficacy of PhotonSplat in handling challenging high-speed camera capture scenarios.
\vspace{-1mm}
\subsection{View-Consistent Colorization}

Next, we evaluate the capabilities of our framework to encode the visual appearance of a scene. 
Given a reference image, we can naively use them as an input condition to pre-trained 2D models to color the grayscale images rendered from our model. 
Specifically we take two pretrained models (1) image based~\cite{kang2023ddcolor} and (2) video-based~\cite{yang2025colormnet}.
We take the rendered views and measure consistency across multiple viewpoints by warping each view with respect to the other based on optical flow~\cite{teed2020raft} and then compute the masked RMSE and LPIPS score~\cite{zhang2018perceptual} across nearby views (short-range consistency) and far-away views (long-range consistency). 
We present quantitative results on the simulated captures in Table. \ref{tab:consis} and we can clearly see that our view consistent colorization module outperforms all baselines. 
Fig. \ref{fig:refcolor} visualizes these results on the real captures, and once again, we can clearly see that our simple 3D aware colorization module outperforms large generative models both in terms of closeness to reference image color and consistency. 

Finally, we compare similar results in the case of no reference image, where PhotonSplat relies on pre-trained 2D models to appropriately encode scene color in a 3D-aware manner. Since we are probabilistically predicting the possible scene color, we do not report quantitative results and instead showcase qualitative comparison against two baseline pre-trained image models DDColor~\cite{kang2023ddcolor} and Deolidy~\cite{salmona2022deoldify} in real captures in Fig. \ref{fig:norefcolor}. We apply these models on the denoised outputs of renders from our method. We see that our view-consistent colorization pipeline can successfully recover accurate and consistent color across multiple viewpoints. Note that in our colorization experiment, we compare only against our method with 2D colorization modules, as naive NANerf and DeblurGS fail to reconstruct geometry accurately(see Fig.~\ref{fig:recon}, and applying them for colorization would further degrade quality. Moreover, this experiment aims to demonstrate the consistent colorization capability of our method over 2D-based approaches.

\begin{figure}[h!]
  \includegraphics[width=0.48\textwidth]{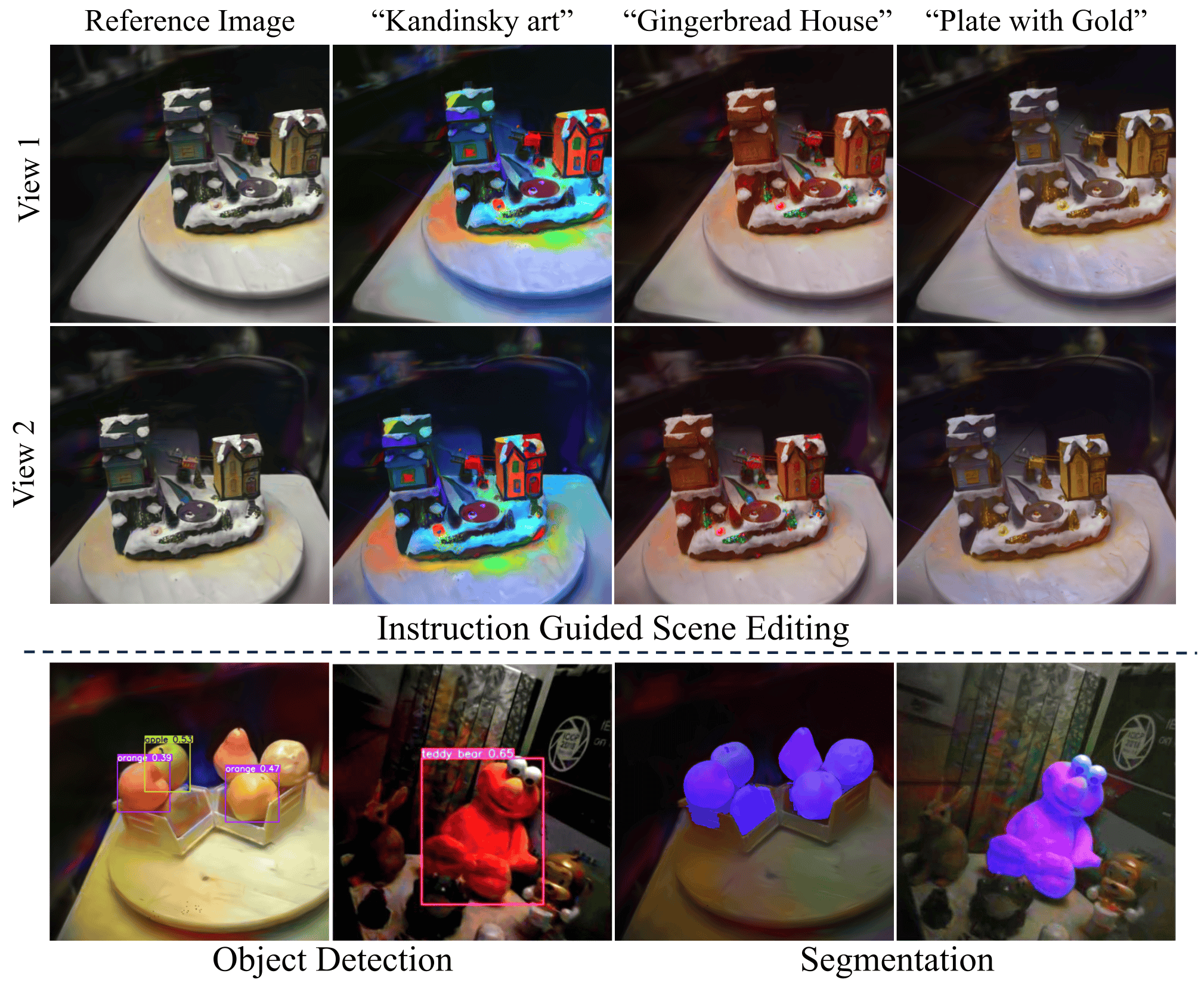}
  \caption{We showcase applications of our rendered color views for instruction guided scene editing (row 1 and 2) and object recognition (row 3). Accurate results indicate that the rendered images are photorealistic and suitable for several downstream tasks.}
  \label{fig:down}
\end{figure}
\vspace{-1.5mm}
\subsection{Downstream Tasks}
Since we can successfully recover the 3D scene and encode visual appearance, it makes it suitable for several downstream tasks. 
We showcase two tasks - instruction-guided scene editing and object recognition on top of a fully optimized model for a given scene. 
In the case of instruction-guided scene editing, we apply InstructPix2Pix~\cite{brooks2023instructpix2pix} on the reference view and propagate it in a 3D consistent manner using our colorization module. 
For object recognition, we simply run the pretrained detector on each rendered view and visualize outputs. 
In Fig. \ref{fig:down}, we showcase qualitative evidence that the rendered results from PhotonSplat are suitable for both of these tasks. 
This further underscores that the rendered results are realistic and can be operated upon by any existing pre-trained 2D model. 

% \textbf{Appearance Editing: } Our consistent colorization approach enables applications like text-based editing and style-transfer. After colorizing the scene, a single frame is processed using a diffusion-based model, such as instruct-pix2pix, which applies text-based appearance changes while preserving the scene's geometry. The results on a real-world dataset with various text prompts are shown in Fig. \ref{}. Our method effectively modifies the texture of the scene while maintaining color consistency across viewpoints.
\vspace{-1mm}
\subsection{Dynamic scenes}
\begin{figure}[t!]
  \includegraphics[width=0.48\textwidth]{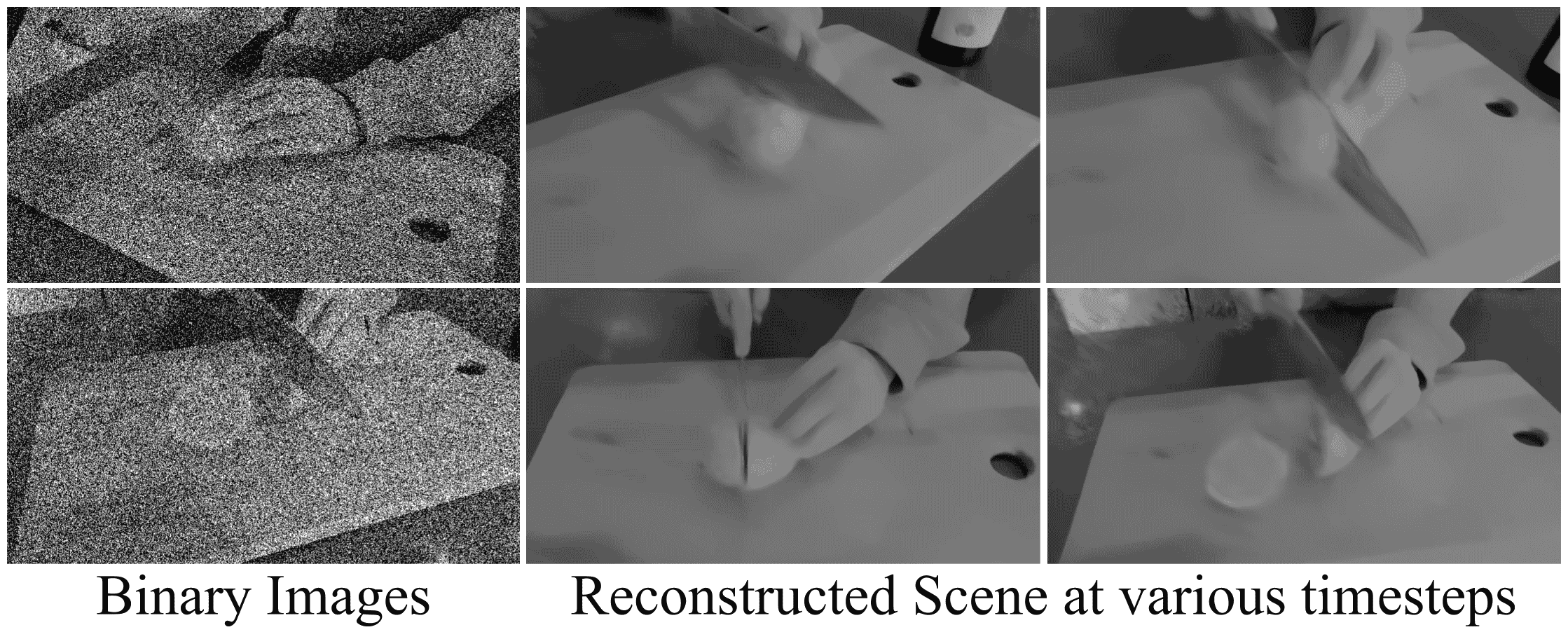}
\caption{Our method reconstructs the 4D dynamic scenes from multi-view binary images. We showcase novel-view renderings across different viewpoints and timestamps. }

  \label{fig:dyn}
  % \vspace{-3em}
\end{figure}

All the results discussed above consider the case of static scenes, but with high-speed camera motion, which simulates motion blur. 
Motion blur can also arise from objects moving in the scene. 
Our photon modelling framework can be easily incorporated into any reconstruction technique that leverages a splat-based scene representation. 
To evaluate the potential use case on 4D data, we extend our technique onto ~\cite{wu20244d}, which additionally models Gaussian deformations dependent on time to effectively model dynamic scenes. 
As described in Sec. \ref{sec:data}, we simulate SPAD-like image captures on the HyperNeRF dataset~\cite{park2021hypernerf}. Fig. \ref{fig:dyn} shows that we can successfully recover the 3D scene at different time intervals. 

% \textbf{Dynamic Scenes: } Real-world multi-view captures often involve dynamic scenes with object motion. While our framework primarily supports static scenes, we extend it to 4D dynamic scenes using the methodology of \cite{wu20244d}, which models Gaussian deformations based on timestamps. We evaluate our method on both synthetic and real-world datasets, demonstrating results in Fig. \ref{}. The model successfully captures scene dynamics and renders novel views consistently across different timestamps and viewpoints.

\vspace{-1mm}
\subsection{Ablation Study}
\label{sec:ablation}
\begin{figure}[t!]
  \includegraphics[width=0.48\textwidth]{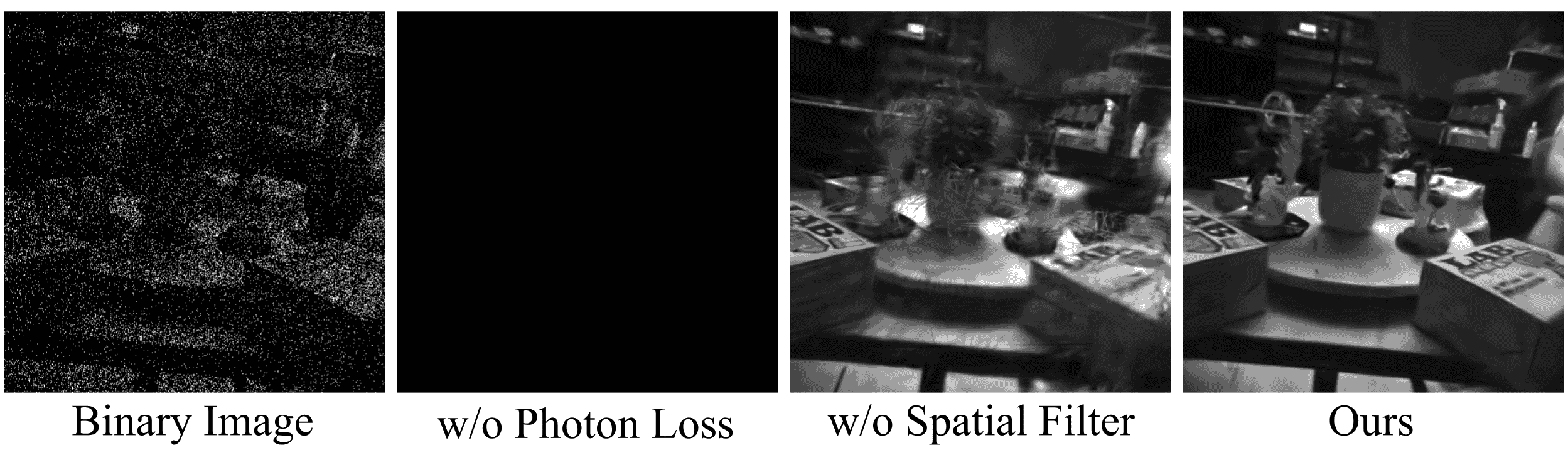}
\caption{\textbf{Ablation Studies:} We demonstrate that each proposed component is crucial for reducing artifacts and inconsistencies, leading to high quality renderings.}
  \label{fig:abl}
  % \vspace{-3em}
\end{figure}

\begin{figure}[t!]
  \includegraphics[width=0.48\textwidth]{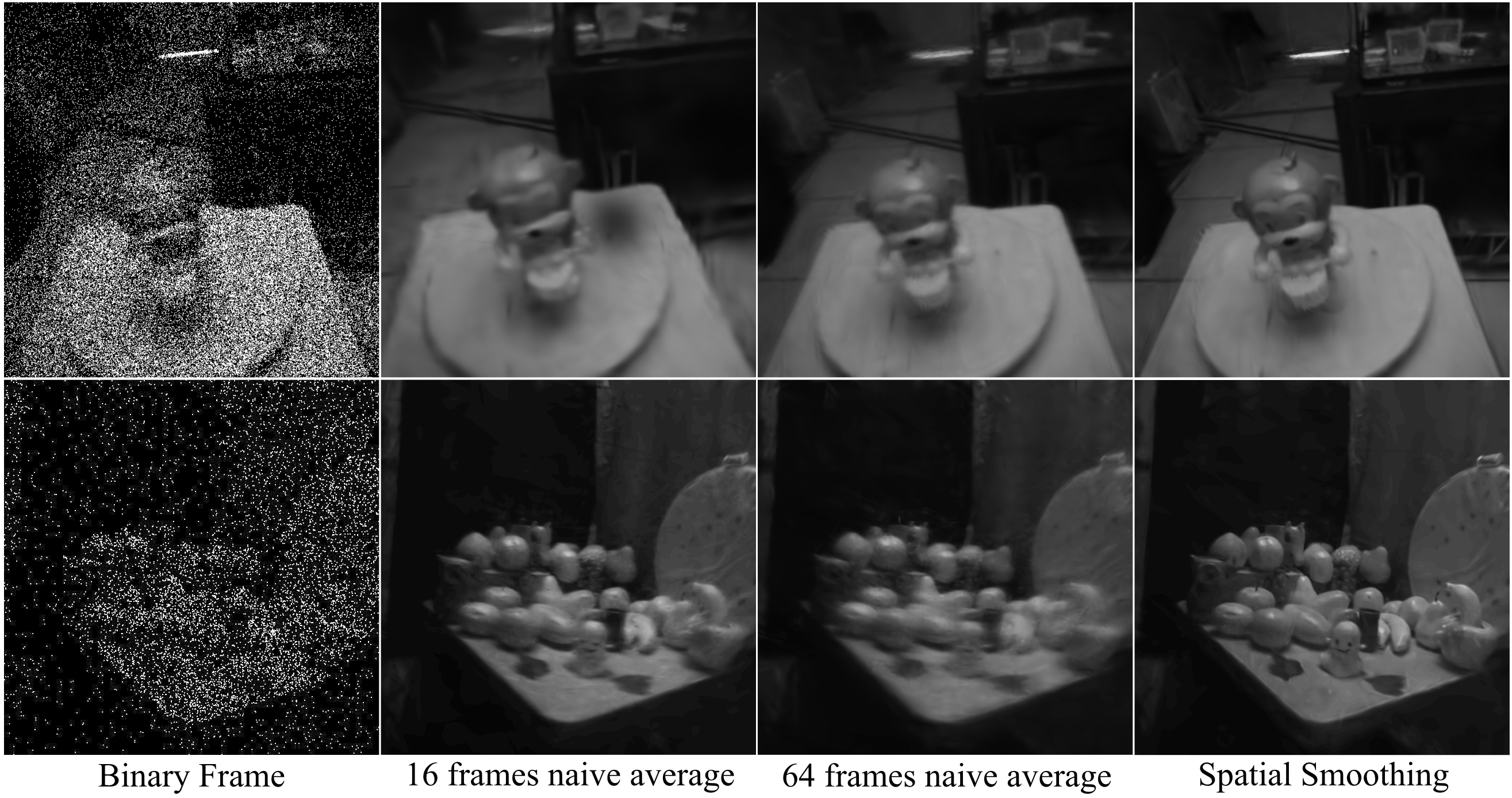}
\caption{\textbf{Ablation Studies:} Comparing our spatial smoothing method with naive frame averaging.}
  \label{fig:abl_spatial}
  % \vspace{-3em}
\end{figure}

% To evaluate the individual components of our proposed framework, we perform several ablations and visualize results in Fig. \ref{fig:abl}. We first train a vanilla Gaussian Splatting model on the original SPAD images, which collapses the entire model to almost black renderings. Removing the smoothening regularization yields significant artifacts in the rendered images, while our entire pipeline yields the best results. 
\textbf{Key Components:} To evaluate the individual components of our proposed framework, we perform several ablations and visualize results in Fig. \ref{fig:abl}. All models are trained as described in Sec. \ref{sec:implementation} and evaluated on the task of view synthesis from binary images. We first train a Vanilla-GS model on binary images without modifying the loss function, which collapses the entire model to almost black renderings. Introducing the proposed photon loss enables effective 3D photon modeling and novel view synthesis but introduces noise due to the binary nature of SPAD images. Finally, incorporating a 3D spatial filter reduces this noise, preventing the GS model from overfitting to the noise. 

\noindent \textbf{Spatial Smoothing vs Averaging Frames:} Our method requires no scene-specific fine-tuning, as the per-frame regularization during rendering is independent of camera motion, and the same $\sigma$ value generalizes well across all scenes. In contrast, methods relying on averaged frames require scene-specific tuning, as the optimal number of frames varies by scene depending on camera speed. The ablation study in the above figure compares naive frame averaging with spatial smoothing across two scenes, highlighting that while averaging needs scene-specific adjustment, our approach performs better and remains scene-agnostic.

\label{sec:experiments}

\begin{figure}[t!]
  \includegraphics[width=0.48\textwidth]{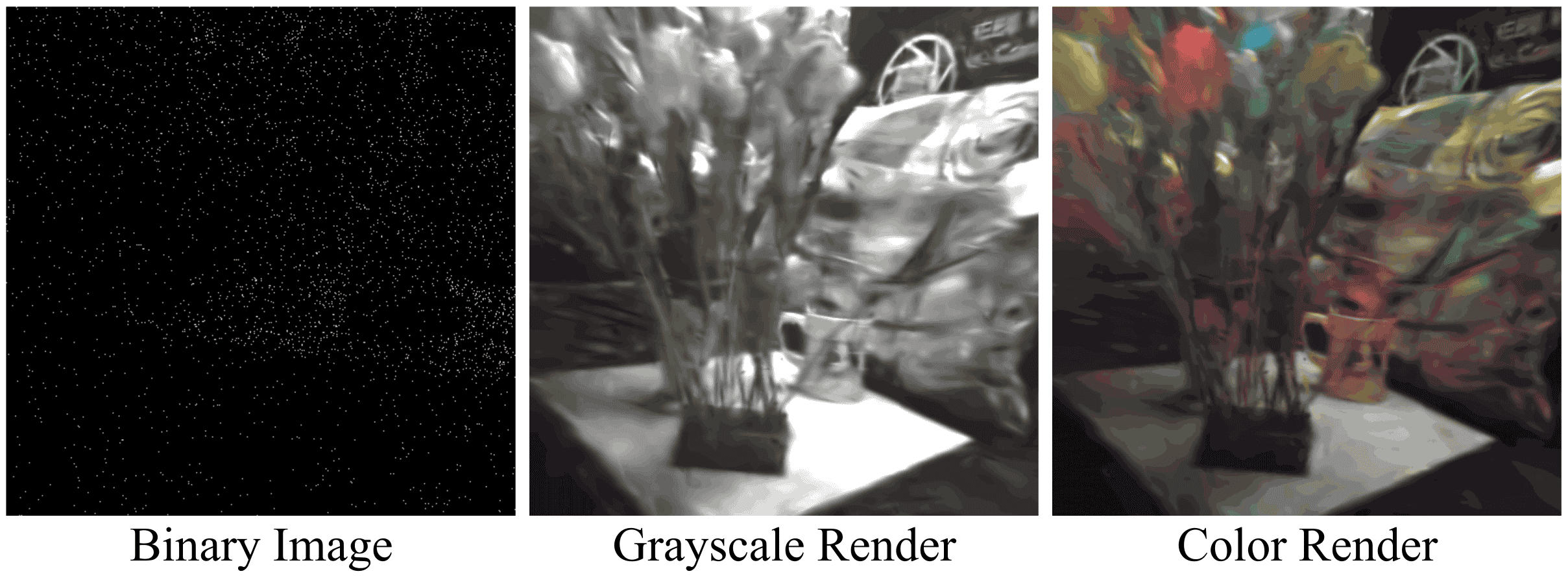}
\caption{Our method fails to reconstruct images in extreme low-light due to insufficient information in the input SPAD images.}
  \label{fig:fail}
  % \vspace{-3em}
\end{figure}

% \vspace{-2mm}

\section{Discussion}
% \section{Conclusion}
\label{sec:conclusion}

%\textbf{Conclusion. }
We introduce \textit{PhotonSplat}, a novel framework that reconstructs a 3D scene from multi-view SPAD captures obtained from a high-speed camera setup. 
By explicitly modeling the photon hitting probabilities into the gaussian splatting framework, along with suitable regularization, we can effectively recover the scene geometry. 
To ensure practicality of our method, we also propose a 3D aware colorization module to encode visual appearance into the reconstructed scene, that is critical for several downstream applications. Furthermore, we extend our framework to enable the reconstruction of dynamic 4D scenes. We conduct several experiments to evaluate our proposed technique, notably even on real SPAD captures. 
We hope our work paves the way for the possibility of leveraging other camera sensing modalities for 3D reconstruction. 

% integrates SPAD imagery into the Gaussian splatting approach for 3D scene reconstruction from high-speed moving cameras. Our method explicitly models pixel-wise photon-hitting probabilities using a binary cross-entropy loss and effectively reduces noise in binary images through a 3D spatial filter. Additionally, our framework supports view-consistent colorization, a critical feature for downstream applications. Notably, PhotonSplat is the first framework to extensively test on real-world multi-view dataset for SPAD Imaging. This work highlights the potential of integrating novel sensors into 3D reconstruction frameworks, expanding the possibilities for addressing challenging real-world scenarios.
\vspace{1mm}
\noindent \textbf{Limitations and Future Work. }When the photon detections from the SPAD sensor are poor e.g. in the case of extreme low-light, the reconstructed scenes lose quality. Recovering poses from such images are also extremely difficult, and learning the camera poses simultaneously during training could be a potential solution. Since our framework relies on COLMAP for Structure from Motion (SfM), its inability to handle larger numbers of frames becomes a bottleneck. Future works could build on integrating pose information from the IMU chip to remove the dependency on SfM and preprocessing in real-time applications. We could also explore spline-based pose interpolation to generate intermediate poses and extend our framework to handle higher frame rate. Our method is also compatible with recent SPAD sensors equipped with color filters, enabling support for color binary frames. It naturally extends to all three color channels, and since each channel remains single-bit, our photon modeling and loss formulation remain applicable, ensuring compatibility with future hardware.

% Despite its strong performance on real-world and synthetic datasets, our framework has some limitations. 
% It struggles in extremely low-light conditions where insufficient photons hit the sensor, as shown in Fig. \ref{fig:fail}. 
% The lack of sufficient information makes it challenging for the model to estimate photon probabilities and recover the scene accurately. 
% Enhancing the framework to operate effectively under these conditions could have significant real-world applications. 
% Additionally, when averaging frames for Structure-from-Motion (SfM), large camera movements can introduce significant motion blur, leading to pose estimation failures. 
% Jointly optimizing camera poses and 3D Gaussians in an end-to-end manner could potentially resolve this issue and could be another relevant direction in the future.  

% \newpage
% \clearpage

\bibliographystyle{IEEEtran}
\bibliography{references}

\begin{thebibliography}{10}
\providecommand{\url}[1]{#1}
\csname url@samestyle\endcsname
\providecommand{\newblock}{\relax}
\providecommand{\bibinfo}[2]{#2}
\providecommand{\BIBentrySTDinterwordspacing}{\spaceskip=0pt\relax}
\providecommand{\BIBentryALTinterwordstretchfactor}{4}
\providecommand{\BIBentryALTinterwordspacing}{\spaceskip=\fontdimen2\font plus
\BIBentryALTinterwordstretchfactor\fontdimen3\font minus \fontdimen4\font\relax}
\providecommand{\BIBforeignlanguage}[2]{{%
\expandafter\ifx\csname l@#1\endcsname\relax
\typeout{** WARNING: IEEEtran.bst: No hyphenation pattern has been}%
\typeout{** loaded for the language `#1'. Using the pattern for}%
\typeout{** the default language instead.}%
\else
\language=\csname l@#1\endcsname
\fi
#2}}
\providecommand{\BIBdecl}{\relax}
\BIBdecl

\bibitem{mildenhall2021nerf}
B.~Mildenhall, P.~P. Srinivasan, M.~Tancik, J.~T. Barron, R.~Ramamoorthi, and R.~Ng, ``Nerf: Representing scenes as neural radiance fields for view synthesis,'' \emph{Communications of the ACM}, vol.~65, no.~1, pp. 99--106, 2021.

\bibitem{kerbl3Dgaussians}
\BIBentryALTinterwordspacing
B.~Kerbl, G.~Kopanas, T.~Leimk{\"u}hler, and G.~Drettakis, ``3d gaussian splatting for real-time radiance field rendering,'' \emph{ACM Transactions on Graphics}, vol.~42, no.~4, July 2023. [Online]. Available: \url{https://repo-sam.inria.fr/fungraph/3d-gaussian-splatting/}
\BIBentrySTDinterwordspacing

\bibitem{8449092}
A.~C. Ulku, C.~Bruschini, I.~M. Antolović, Y.~Kuo, R.~Ankri, S.~Weiss, X.~Michalet, and E.~Charbon, ``A 512 × 512 spad image sensor with integrated gating for widefield flim,'' \emph{IEEE Journal of Selected Topics in Quantum Electronics}, vol.~25, no.~1, pp. 1--12, 2019.

\bibitem{Liu_2022_WACV}
Y.~Liu, F.~Gutierrez-Barragan, A.~Ingle, M.~Gupta, and A.~Velten, ``Single-photon camera guided extreme dynamic range imaging,'' in \emph{Proceedings of the IEEE/CVF Winter Conference on Applications of Computer Vision (WACV)}, January 2022, pp. 1575--1585.

\bibitem{s16071122}
\BIBentryALTinterwordspacing
N.~A.~W. Dutton, I.~Gyongy, L.~Parmesan, and R.~K. Henderson, ``Single photon counting performance and noise analysis of cmos spad-based image sensors,'' \emph{Sensors}, vol.~16, no.~7, 2016. [Online]. Available: \url{https://www.mdpi.com/1424-8220/16/7/1122}
\BIBentrySTDinterwordspacing

\bibitem{jungerman2024radiance}
S.~Jungerman and M.~Gupta, ``Radiance fields from photons,'' \emph{arXiv preprint arXiv:2407.09386}, 2024.

\bibitem{kang2023ddcolor}
X.~Kang, T.~Yang, W.~Ouyang, P.~Ren, L.~Li, and X.~Xie, ``Ddcolor: Towards photo-realistic image colorization via dual decoders,'' in \emph{Proceedings of the IEEE/CVF International Conference on Computer Vision}, 2023, pp. 328--338.

\bibitem{ma2023seeing}
S.~Ma, V.~Sundar, P.~Mos, C.~Bruschini, E.~Charbon, and M.~Gupta, ``Seeing photons in color,'' \emph{ACM Transactions on Graphics (TOG)}, vol.~42, no.~4, pp. 1--16, 2023.

\bibitem{gnanasambandam2019megapixel}
A.~Gnanasambandam, O.~Elgendy, J.~Ma, and S.~H. Chan, ``Megapixel photon-counting color imaging using quanta image sensor,'' \emph{Optics express}, vol.~27, no.~12, pp. 17\,298--17\,310, 2019.

\bibitem{purohit2024generative}
V.~Purohit, J.~Luo, Y.~Chi, Q.~Guo, S.~H. Chan, and Q.~Qiu, ``Generative quanta color imaging,'' in \emph{Proceedings of the IEEE/CVF Conference on Computer Vision and Pattern Recognition}, 2024, pp. 25\,138--25\,148.

\bibitem{Kirmani2014-ey}
A.~Kirmani, D.~Venkatraman, D.~Shin, A.~Cola{\c c}o, F.~N.~C. Wong, J.~H. Shapiro, and V.~K. Goyal, ``\BIBforeignlanguage{en}{First-photon imaging},'' \emph{\BIBforeignlanguage{en}{Science}}, vol. 343, no. 6166, pp. 58--61, Jan. 2014.

\bibitem{Shin2016-du}
D.~Shin, F.~Xu, D.~Venkatraman, R.~Lussana, F.~Villa, F.~Zappa, V.~K. Goyal, F.~N.~C. Wong, and J.~H. Shapiro, ``\BIBforeignlanguage{en}{Photon-efficient imaging with a single-photon camera},'' \emph{\BIBforeignlanguage{en}{Nat. Commun.}}, vol.~7, no.~1, p. 12046, Jun. 2016.

\bibitem{ingle2019high}
A.~Ingle, A.~Velten, and M.~Gupta, ``High flux passive imaging with single-photon sensors,'' in \emph{Proceedings of the IEEE/CVF Conference on Computer Vision and Pattern Recognition}, 2019, pp. 6760--6769.

\bibitem{liu2022single}
Y.~Liu, F.~Gutierrez-Barragan, A.~Ingle, M.~Gupta, and A.~Velten, ``Single-photon camera guided extreme dynamic range imaging,'' in \emph{Proceedings of the IEEE/CVF Winter Conference on Applications of Computer Vision}, 2022, pp. 1575--1585.

\bibitem{goyal2021photon}
B.~Goyal and M.~Gupta, ``Photon-starved scene inference using single photon cameras,'' in \emph{Proceedings of the IEEE/CVF International Conference on Computer Vision}, 2021, pp. 2512--2521.

\bibitem{Buttafava:15}
\BIBentryALTinterwordspacing
M.~Buttafava, J.~Zeman, A.~Tosi, K.~Eliceiri, and A.~Velten, ``Non-line-of-sight imaging using a time-gated single photon avalanche diode,'' \emph{Opt. Express}, vol.~23, no.~16, pp. 20\,997--21\,011, Aug 2015. [Online]. Available: \url{https://opg.optica.org/oe/abstract.cfm?URI=oe-23-16-20997}
\BIBentrySTDinterwordspacing

\bibitem{OToole2018-bk}
O'Toole, Matthew, Lindell, D.~B, and G.~Wetzstein, ``\BIBforeignlanguage{en}{Confocal non-line-of-sight imaging based on the light-cone transform},'' \emph{\BIBforeignlanguage{en}{Nature}}, vol. 555, no. 7696, pp. 338--341, Mar. 2018.

\bibitem{Callenberg2021CheapSPAD}
C.~Callenberg, Z.~Shi, F.~Heide, and M.~B. Hullin, ``Low-cost spad sensing for non-line-of-sight tracking, material classification and depth imaging,'' \emph{ACM Trans. Graph. (SIGGRAPH)}, vol.~40, no.~4, 2021.

\bibitem{jungerman2023panoramas}
S.~Jungerman, A.~Ingle, and M.~Gupta, ``Panoramas from photons,'' \emph{arXiv preprint arXiv:2309.03811}, 2023.

\bibitem{ma2020quanta}
S.~Ma, S.~Gupta, A.~C. Ulku, C.~Bruschini, E.~Charbon, and M.~Gupta, ``Quanta burst photography,'' \emph{ACM Transactions on Graphics (TOG)}, vol.~39, no.~4, pp. 79--1, 2020.

\bibitem{mildenhall2020nerf}
B.~Mildenhall, P.~P. Srinivasan, M.~Tancik, J.~T. Barron, R.~Ramamoorthi, and R.~Ng, ``Nerf: Representing scenes as neural radiance fields for view synthesis,'' in \emph{ECCV}, 2020.

\bibitem{ma2022deblurnerf}
L.~Ma, X.~Li, J.~Liao, Q.~Zhang, X.~Wang, J.~Wang, and P.~V. Sander, ``Deblur-nerf: Neural radiance fields from blurry images,'' 2022.

\bibitem{wang2023badnerf}
P.~Wang, L.~Zhao, R.~Ma, and P.~Liu, ``Bad-nerf: Bundle adjusted deblur neural radiance fields,'' 2023.

\bibitem{peng2023pdrf}
C.~Peng and R.~Chellappa, ``Pdrf: progressively deblurring radiance field for fast scene reconstruction from blurry images,'' in \emph{Proceedings of the AAAI Conference on Artificial Intelligence}, vol.~37, no.~2, 2023, pp. 2029--2037.

\bibitem{cui2024aleth}
Z.~Cui, L.~Gu, X.~Sun, X.~Ma, Y.~Qiao, and T.~Harada, ``Aleth-nerf: Illumination adaptive nerf with concealing field assumption,'' in \emph{Proceedings of the AAAI Conference on Artificial Intelligence}, vol.~38, no.~2, 2024, pp. 1435--1444.

\bibitem{wang2023lighting}
H.~Wang, X.~Xu, K.~Xu, and R.~W. Lau, ``Lighting up nerf via unsupervised decomposition and enhancement,'' in \emph{Proceedings of the IEEE/CVF International Conference on Computer Vision}, 2023, pp. 12\,632--12\,641.

\bibitem{wu20244d}
G.~Wu, T.~Yi, J.~Fang, L.~Xie, X.~Zhang, W.~Wei, W.~Liu, Q.~Tian, and X.~Wang, ``4d gaussian splatting for real-time dynamic scene rendering,'' in \emph{Proceedings of the IEEE/CVF Conference on Computer Vision and Pattern Recognition}, 2024, pp. 20\,310--20\,320.

\bibitem{duan20244d}
Y.~Duan, F.~Wei, Q.~Dai, Y.~He, W.~Chen, and B.~Chen, ``4d-rotor gaussian splatting: towards efficient novel view synthesis for dynamic scenes,'' in \emph{ACM SIGGRAPH 2024 Conference Papers}, 2024, pp. 1--11.

\bibitem{guedon2024sugar}
A.~Gu{\'e}don and V.~Lepetit, ``Sugar: Surface-aligned gaussian splatting for efficient 3d mesh reconstruction and high-quality mesh rendering,'' in \emph{Proceedings of the IEEE/CVF Conference on Computer Vision and Pattern Recognition}, 2024, pp. 5354--5363.

\bibitem{huang20242d}
B.~Huang, Z.~Yu, A.~Chen, A.~Geiger, and S.~Gao, ``2d gaussian splatting for geometrically accurate radiance fields,'' in \emph{ACM SIGGRAPH 2024 Conference Papers}, 2024, pp. 1--11.

\bibitem{chen2024deblur}
W.~Chen and L.~Liu, ``Deblur-gs: 3d gaussian splatting from camera motion blurred images,'' \emph{Proceedings of the ACM on Computer Graphics and Interactive Techniques}, vol.~7, no.~1, pp. 1--15, 2024.

\bibitem{lee2024deblurring}
B.~Lee, H.~Lee, X.~Sun, U.~Ali, and E.~Park, ``Deblurring 3d gaussian splatting,'' \emph{arXiv preprint arXiv:2401.00834}, 2024.

\bibitem{ye2024thermal}
T.~Ye, Q.~Wu, J.~Deng, G.~Liu, L.~Liu, S.~Xia, L.~Pang, W.~Yu, and L.~Pei, ``Thermal-nerf: Neural radiance fields from an infrared camera,'' \emph{arXiv preprint arXiv:2403.10340}, 2024.

\bibitem{lin2024thermalnerf}
Y.~Y. Lin, X.-Y. Pan, S.~Fridovich-Keil, and G.~Wetzstein, ``Thermalnerf: Thermal radiance fields,'' in \emph{2024 IEEE International Conference on Computational Photography (ICCP)}.\hskip 1em plus 0.5em minus 0.4em\relax IEEE, 2024, pp. 1--12.

\bibitem{lu2024thermalgaussian}
R.~Lu, H.~Chen, Z.~Zhu, Y.~Qin, M.~Lu, L.~Zhang, C.~Yan, and A.~Xue, ``Thermalgaussian: Thermal 3d gaussian splatting,'' \emph{arXiv preprint arXiv:2409.07200}, 2024.

\bibitem{xiong2024event3dgs}
T.~Xiong, J.~Wu, B.~He, C.~Fermuller, Y.~Aloimonos, H.~Huang, and C.~Metzler, ``Event3dgs: Event-based 3d gaussian splatting for high-speed robot egomotion,'' in \emph{8th Annual Conference on Robot Learning}, 2024.

\bibitem{deguchi2024e2gs}
H.~Deguchi, M.~Masuda, T.~Nakabayashi, and H.~Saito, ``E2gs: Event enhanced gaussian splatting,'' in \emph{2024 IEEE International Conference on Image Processing (ICIP)}.\hskip 1em plus 0.5em minus 0.4em\relax IEEE, 2024, pp. 1676--1682.

\bibitem{madavan2024ganesh}
R.~R. Madavan, A.~Kaimal, B.~KV, V.~Gupta, R.~Choudhary, C.~Shanmuganathan, and K.~Mitra, ``Ganesh: Generalizable nerf for lensless imaging,'' \emph{arXiv preprint arXiv:2411.04810}, 2024.

\bibitem{wang2018high}
T.-C. Wang, M.-Y. Liu, J.-Y. Zhu, A.~Tao, J.~Kautz, and B.~Catanzaro, ``High-resolution image synthesis and semantic manipulation with conditional gans,'' in \emph{Proceedings of the IEEE conference on computer vision and pattern recognition}, 2018, pp. 8798--8807.

\bibitem{spadcamera}
{SPAD Camera}, ``Spad camera,'' \url{https://piimaging.com/product-spad512s}, 2021.

\bibitem{fu2023colmap}
Y.~Fu, S.~Liu, A.~Kulkarni, J.~Kautz, A.~A. Efros, and X.~Wang, ``Colmap-free 3d gaussian splatting,'' \emph{arXiv preprint arXiv:2312.07504}, 2023.

\bibitem{pan2024glomap}
L.~Pan, D.~Barath, M.~Pollefeys, and J.~L. Sch\"{o}nberger, ``{Global Structure-from-Motion Revisited},'' in \emph{European Conference on Computer Vision (ECCV)}, 2024.

\bibitem{pearl2022nan}
N.~Pearl, T.~Treibitz, and S.~Korman, ``Nan: Noise-aware nerfs for burst-denoising,'' in \emph{Proceedings of the IEEE/CVF Conference on Computer Vision and Pattern Recognition}, 2022, pp. 12\,672--12\,681.

\bibitem{yang2025colormnet}
Y.~Yang, J.~Dong, J.~Tang, and J.~Pan, ``Colormnet: A memory-based deep spatial-temporal feature propagation network for video colorization,'' in \emph{European Conference on Computer Vision}.\hskip 1em plus 0.5em minus 0.4em\relax Springer, 2025, pp. 336--352.

\bibitem{teed2020raft}
Z.~Teed and J.~Deng, ``Raft: Recurrent all-pairs field transforms for optical flow,'' in \emph{Computer Vision--ECCV 2020: 16th European Conference, Glasgow, UK, August 23--28, 2020, Proceedings, Part II 16}.\hskip 1em plus 0.5em minus 0.4em\relax Springer, 2020, pp. 402--419.

\bibitem{zhang2018perceptual}
R.~Zhang, P.~Isola, A.~A. Efros, E.~Shechtman, and O.~Wang, ``The unreasonable effectiveness of deep features as a perceptual metric,'' in \emph{CVPR}, 2018.

\bibitem{salmona2022deoldify}
A.~Salmona, L.~Bouza, and J.~Delon, ``Deoldify: A review and implementation of an automatic colorization method,'' \emph{Image Processing On Line}, vol.~12, pp. 347--368, 2022.

\bibitem{brooks2023instructpix2pix}
T.~Brooks, A.~Holynski, and A.~A. Efros, ``Instructpix2pix: Learning to follow image editing instructions,'' in \emph{Proceedings of the IEEE/CVF Conference on Computer Vision and Pattern Recognition}, 2023, pp. 18\,392--18\,402.

\bibitem{park2021hypernerf}
K.~Park, U.~Sinha, P.~Hedman, J.~T. Barron, S.~Bouaziz, D.~B. Goldman, R.~Martin-Brualla, and S.~M. Seitz, ``Hypernerf: A higher-dimensional representation for topologically varying neural radiance fields,'' \emph{ACM Trans. Graph.}, 2021.

\end{thebibliography}

% \ifpeerreview \else
%%%% For the camera ready version, please fill out this
%%%% biography. Your camera ready should be within a 12 page limit
%%%% including acknowledgments, references and biography.

% \includegraphics command within an optional argument. (You could
% create your own custom macro containing the \includegraphics command
% to make things simpler here.)
% \begin{IEEEbiography}[{\includegraphics[width=1in,height=1.25in,clip,keepaspectratio]{mshell}}]{Michael Shell}

% \end{IEEEbiography}
\begin{IEEEbiography}[{\includegraphics[width=1in,height=1.25in,clip,keepaspectratio]{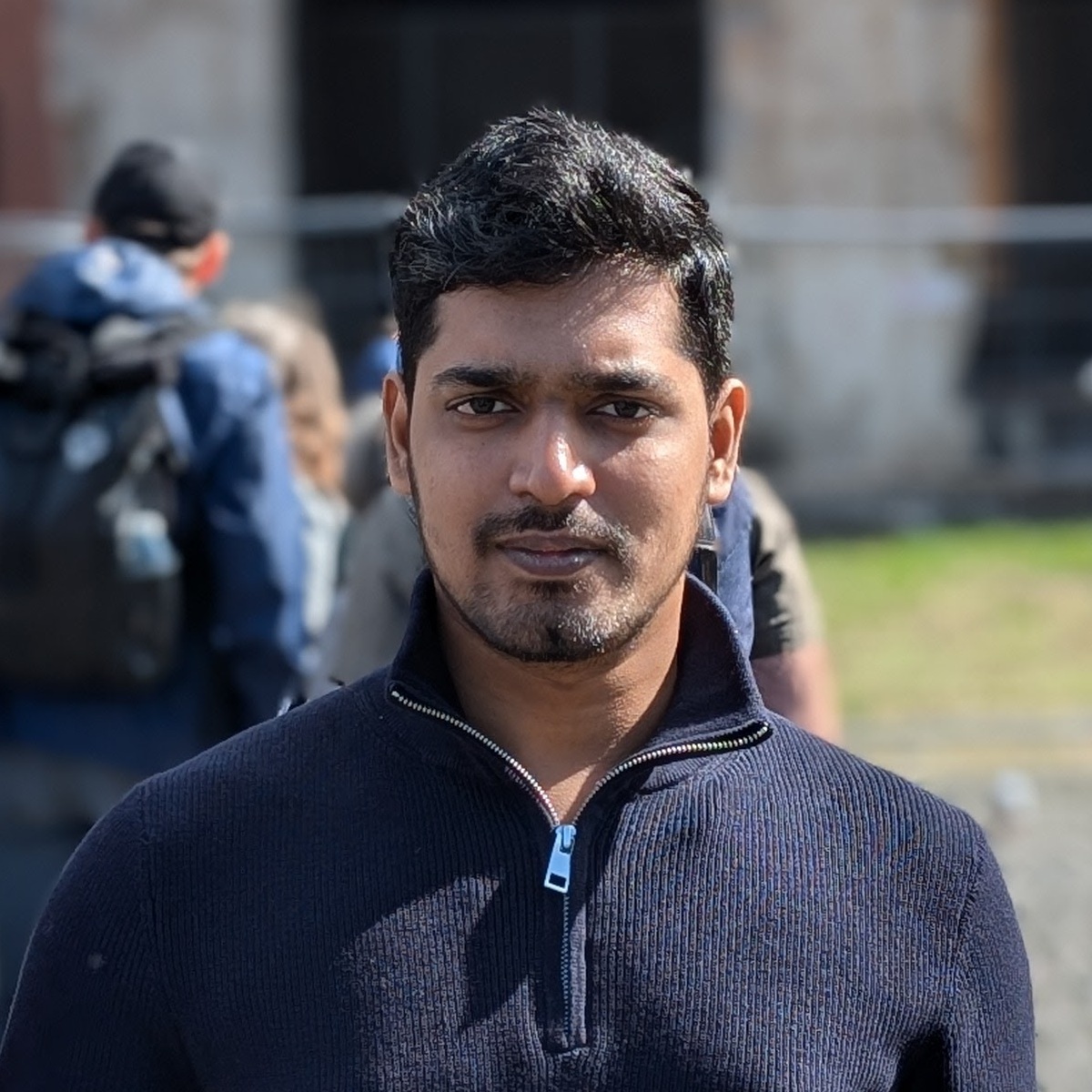}}]{Sai Sri Teja Kuppa}
is a Master's student in Electrical Engineering at IIT Madras advised by Prof Kaushik Mitra. His research interests lie in Computer Vision and Image processing. Teja completed his undergraduate studies at Sastra University. He will be joining Immerso as a full-time researcher this fall 2025. 
\end{IEEEbiography}
\begin{IEEEbiography}[{\includegraphics[width=1in,height=1.25in,clip,keepaspectratio]{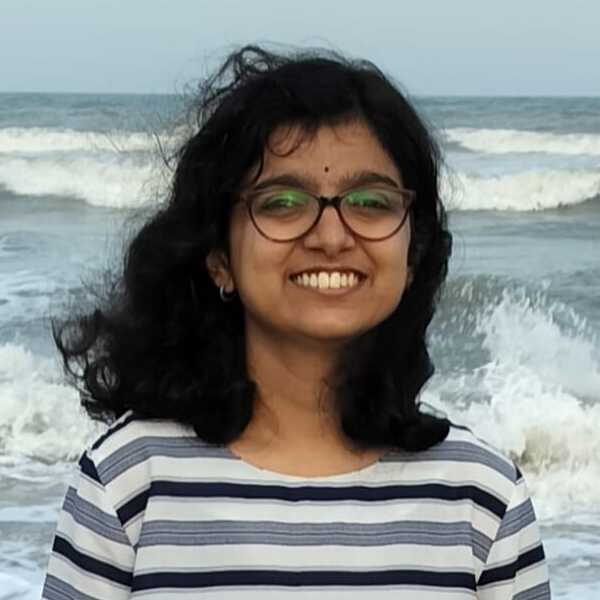}}]{Sreevidya Chintalapati} was a project associate in Computational Imaging Lab, IIT Madras advised by Prof. Kaushik Mitra. She is interested in computational imaging, computer vision and signal processing. Sreevidya completed her undergraduate studies in IIT Gandhinagar majoring in Electrical Engineering. She is an incoming PhD student in Electrical and Computer Engineering at Rice University.

\end{IEEEbiography}
\begin{IEEEbiography}[{\includegraphics[width=1in,height=1.25in,clip,keepaspectratio]{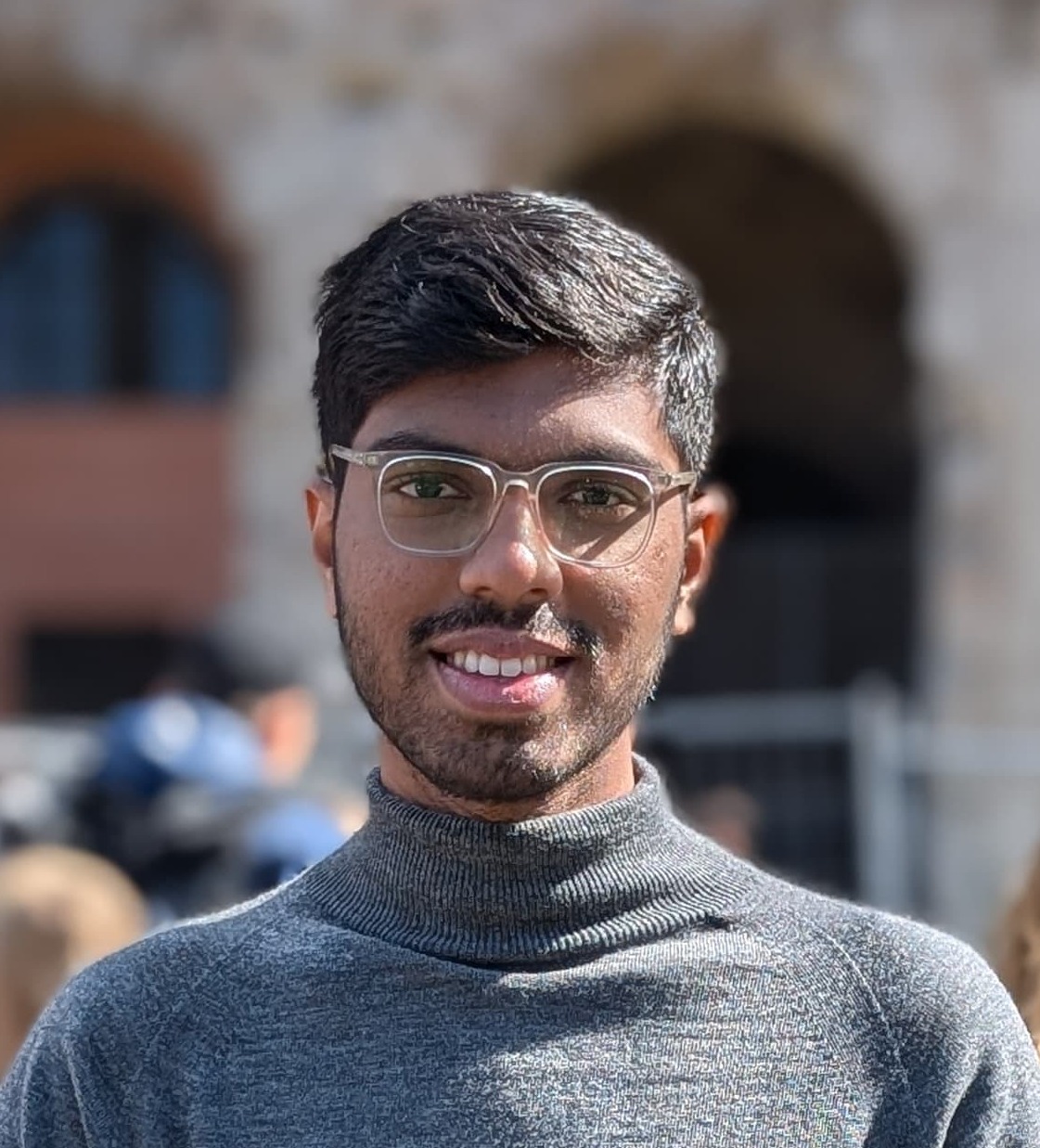}}]{Vinayak Gupta} is a Dual Degree Student at IIT Madras pursuing his BTech in Electrical Engineering and Master's in Data Science. He was part of the Computational Imaging Lab at IIT Madras under the guidance of Prof Kaushik Mitra. His interest lies in Computer Vision and Graphics, especially in 3D/4D reconstruction and generation. He is an incoming PhD student in the Computer Science Department at the University of Maryland, College Park.
\end{IEEEbiography}
\begin{IEEEbiography}[{\includegraphics[width=1in,height=1.25in,clip,keepaspectratio]{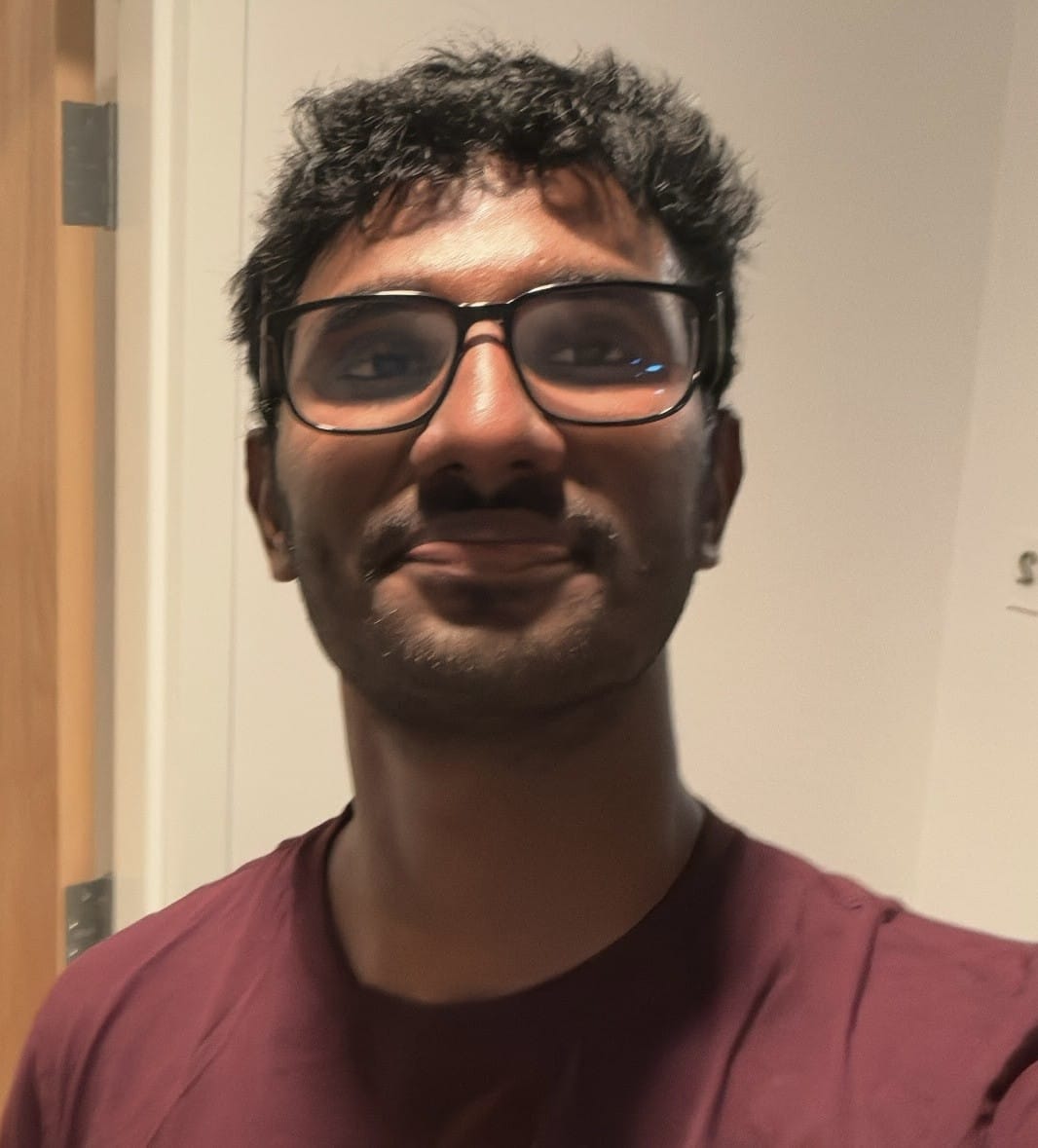}}]{Mukund Varma T}
is a PhD student in Computer Science at the University of California, San Diego advised by Prof Ravi Ramamoorthi and Prof Hao Su.
His research interests lie in the intersection of computer vision, computer graphics, and machine learning, specifically to facilitate high-quality 3D reconstructions and semantic understanding from multiple viewpoints. Mukund completed his Bachelor's and Master's studies at IIT Madras majoring in Mechanical Engineering and Robotics.
\end{IEEEbiography}
\begin{IEEEbiography}[{\includegraphics[width=1in,height=1.25in,clip,keepaspectratio]{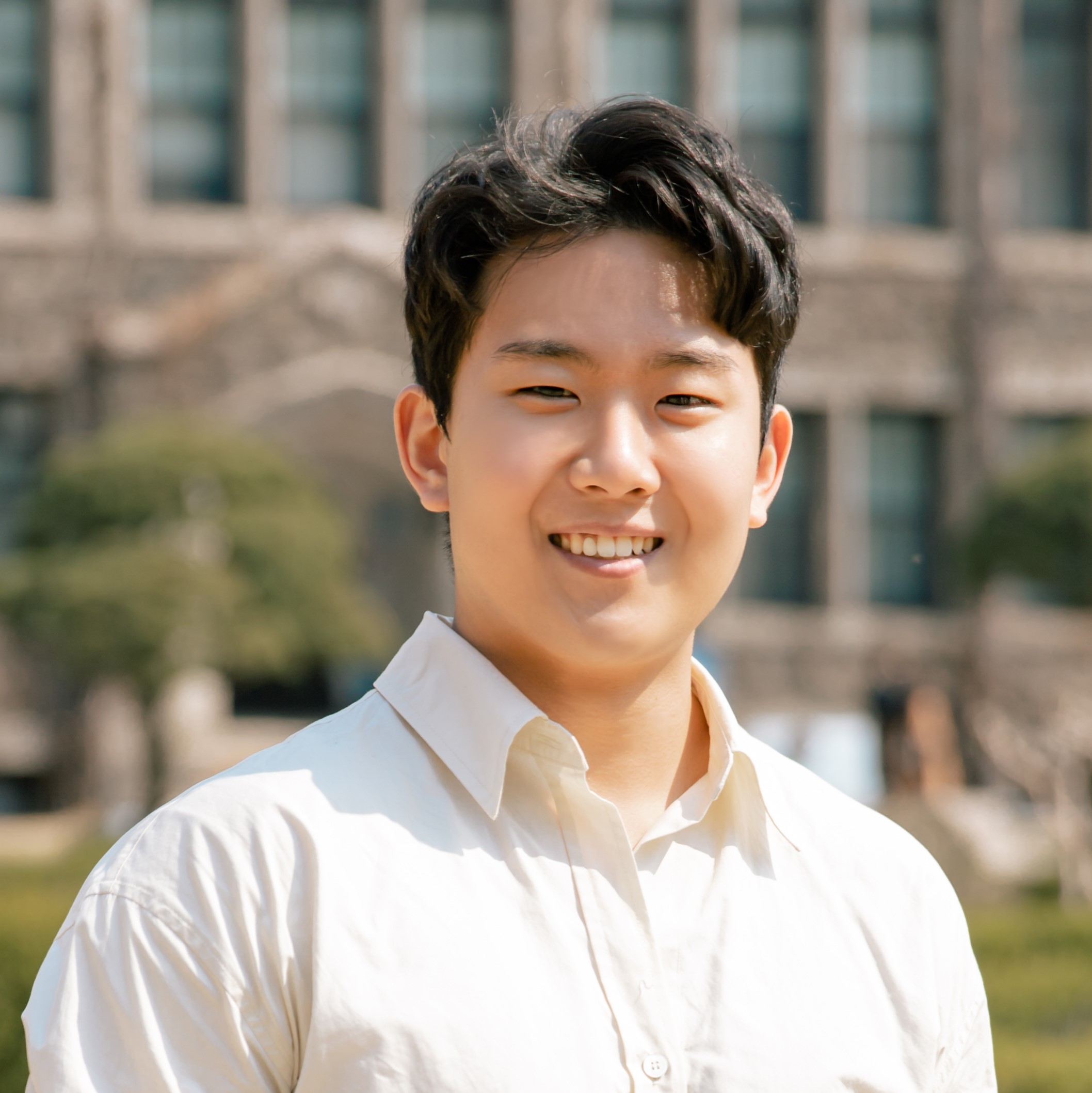}}]{Haejoon Lee}
is a PhD student in Electrical and Computer Engineering at Carnagie Mellon University advised by Prof Aswin C. Sankaranarayanan and Prof Vijayakumar Bhagavatula.
His research interests lie in Computer Vision and Computational Imaging. Haejoon completed his undergraduate studies at Yonsei University, where he majored in Electrical and Electronic Engineering. 
\end{IEEEbiography}
\begin{IEEEbiography}[{\includegraphics[width=1in,height=1.25in,clip,keepaspectratio]{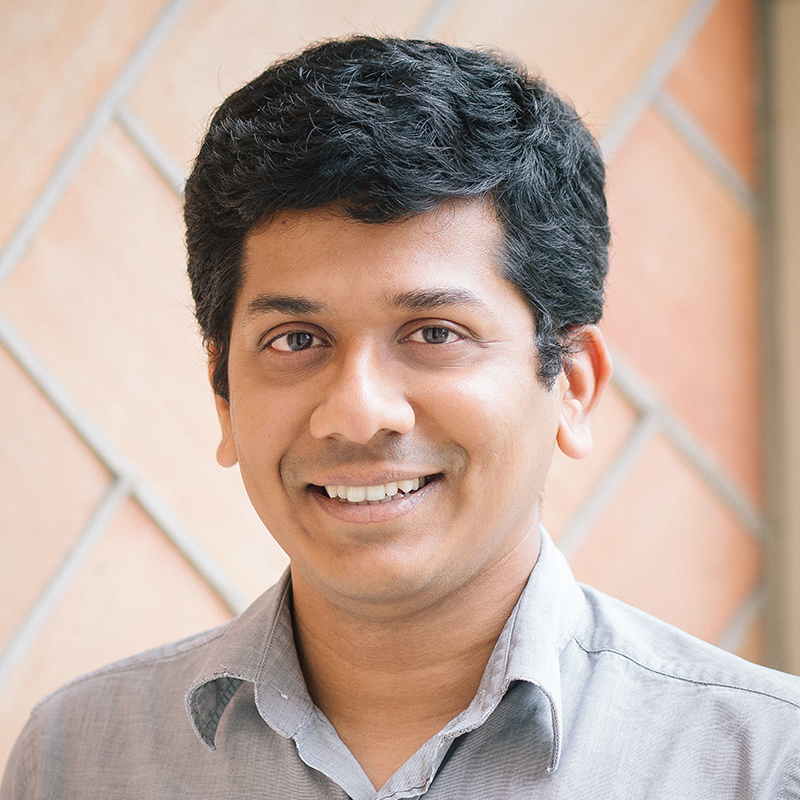}}]{Aswin Sankaranarayanan}
is a Professor of Electrical and Computer Engineering at Carnegie Mellon University and leads the Image Science Lab. His research blends physics-based and learning-based models to co‑design optics, sensors, and inference algorithms for novel computational imaging systems—from displays to non‑line‑of‑sight reconstruction. He earned his Ph.D. from the University of Maryland (2009), where his dissertation received a Distinguished Dissertation Fellowship, and completed postdoctoral work at Rice University. Sankaranarayanan is the recipient of  an NSF CAREER Award (2017), CMU College of Engineering Dean’s Early Career Fellowship (2018), and best paper awards at CVPR 2019, and SIGGRAPH 2023.
\end{IEEEbiography}
\begin{IEEEbiography}[{\includegraphics[width=1in,height=1.25in,clip,keepaspectratio]{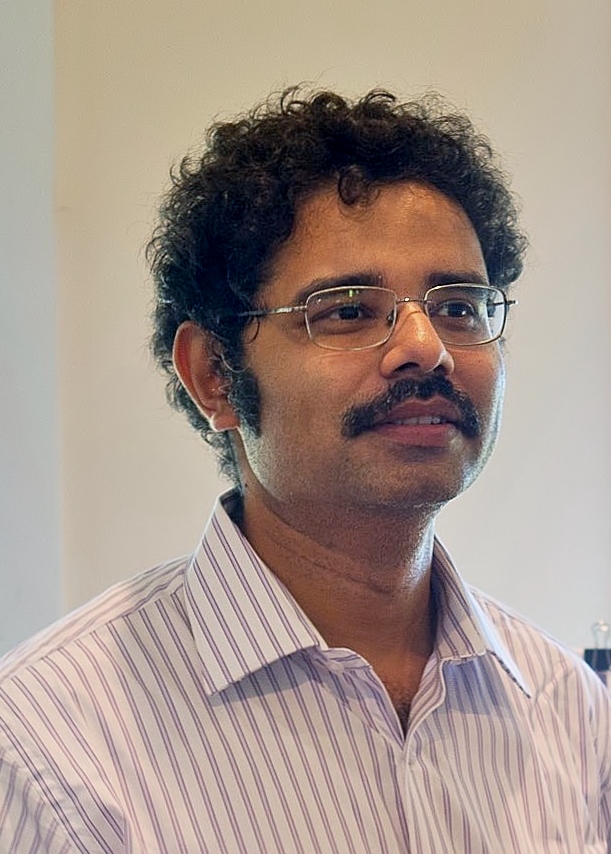}}]{Kaushik Mitra}
is an Associate Professor in the Department of Electrical Engineering at IIT Madras, where he heads the Computational Imaging Lab. Prior to joining IITM, he earned his Ph.D. from the University of Maryland (College Park), focusing on statistical models and optimization for computer vision, followed by postdoctoral research at Rice University. His lab pioneers co‑design of optics and algorithms for computational imaging systems, working on light-field reconstruction, lensless cameras, low‑light imaging, thermal super‑resolution, and deep learning for inverse problems. Prof Kaushik has mentored award‑winning students—most recently four Qualcomm Innovation Fellowship winners (2016, 2017 “super‑winner”, 2020, 2021)—and holds patents such as a “Mini glove-based gesture recognition device”. His team has published in premier venues including ECCV, CVPR, IEEE TIP, TPAMI and ICCV .
\end{IEEEbiography}

%\newpage


%\vfill

% \fi

% \setcounter{page}{1}
% \maketitlesupplementary

% \IEEEtitleabstractindextext{%
% \begin{abstract}
% The abstract goes here. \lipsum[1]
% \input{section/0_abstract} 
% \end{abstract}

% \begin{IEEEkeywords} % Enter keywords here
% \textcolor{red}{Single Photon Cameras, SPADs, Gaussian Splatting, High-Speed Cameras, Computational Imaging}
% \end{IEEEkeywords}
% }

% Make Title
% \ifpeerreview
% \linenumbers \linenumbersep 15pt\relax 
% \author{Paper ID \paperID\IEEEcompsocitemizethanks{\IEEEcompsocthanksitem This paper is under review for ICCP 2025 and the PAMI special issue on computational photography. Do not distribute.}}
% \markboth{Anonymous ICCP 2025 submission ID \paperID}%
% {}
% \fi
% \maketitle
\newpage
\clearpage

\renewcommand{\thesection}{\Alph{section}}
\renewcommand{\thefigure}{\Alph{figure}}
\renewcommand{\thetable}{\Alph{table}}
\setcounter{section}{0}
\setcounter{figure}{0}
\setcounter{table}{0}

\section*{Supplementary Material}
% \renewcommand{\thesection}{\Alph{section}}
% \renewcommand{\thefigure}{\Alph{figure}}
% \renewcommand{\thetable}{\Alph{table}}
% \setcounter{section}{0}
% \setcounter{figure}{0}
% \setcounter{table}{0}
% \appendix
% Sections in Supplementary
% \begin{itemize}
%     \item Demo Video Section
%     \item More Qualitative Results
%     \begin{itemize}
%         \item 3D Recon results on Real Data
%         \item 3D Recon results on Synthetic Data
%         \item Colorisation Results on Real Data
%         \item Colorisation Results on Synthetic Data
%     \end{itemize}
%     \item Details about the baselines
%     \begin{itemize}
%         \item Both 3D Recon and Colorisation
%     \end{itemize}
%     \item Showcase our Real-World Dataset
% \end{itemize}

\section{Demo Video}
\label{sec:supp:demo}

We have provided a project webpage at \href{https://vinayak-vg.github.io/PhotonSplat/}{vinayak-vg.github.io/PhotonSplat/}, including a video demonstration. This video illustrates the versatility and robustness of our framework by presenting a wide range of results on both real-world and synthetic datasets. The video features detailed visual comparisons with baseline methods for 3D reconstruction and colorization tasks to further validate our approach. These comparisons highlight the superior reconstruction quality and consistency achieved by our method, emphasizing its effectiveness across various challenging scenarios.

\section{More Qualitative Results}

\begin{figure*}[h!]
  \includegraphics[width=\textwidth]{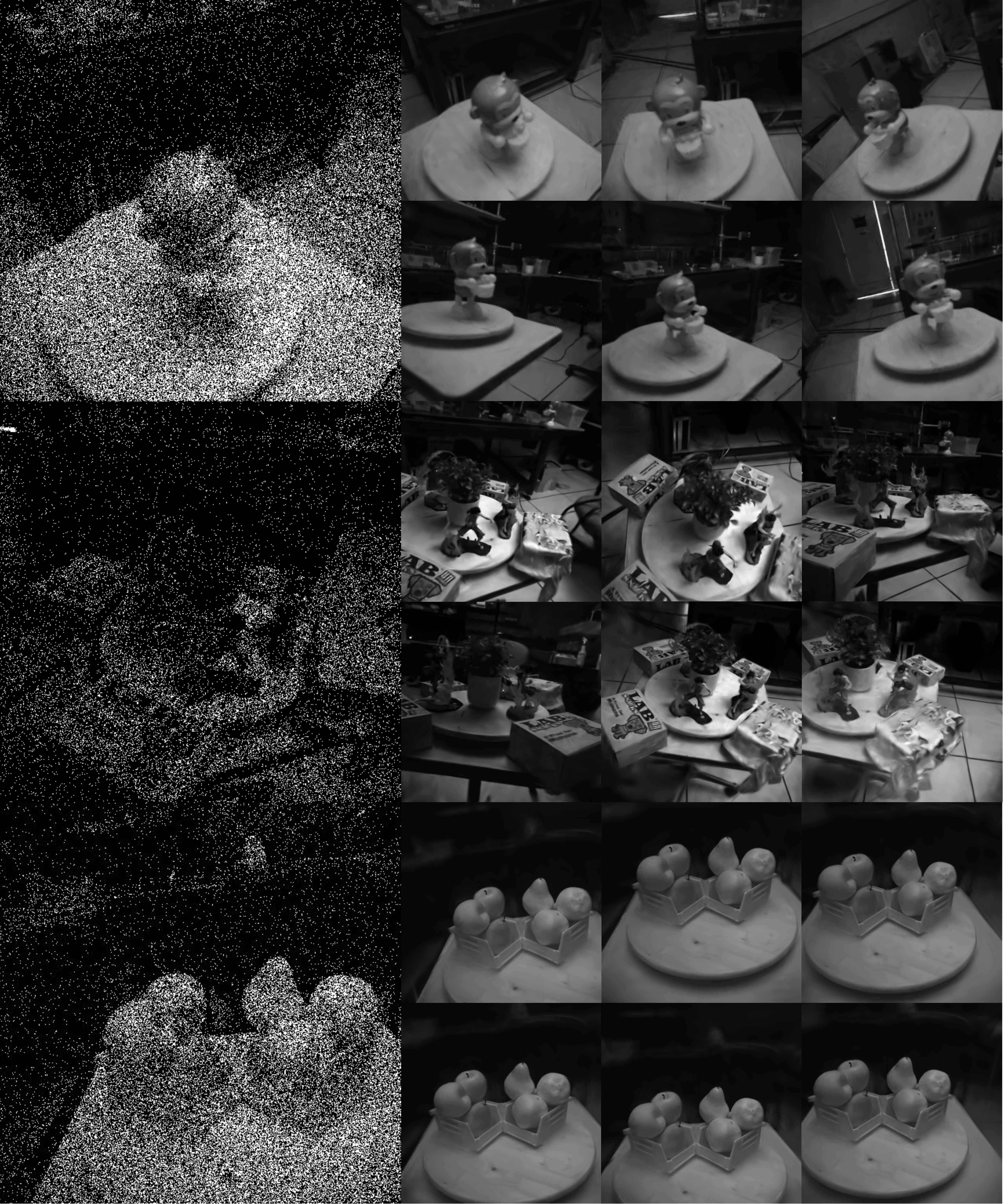}
\caption{3D reconstruction results are presented for real scenes captured using single-photon camera. The first column displays the source binary image, while the subsequent grid showcases rendered novel views. The dataset includes scenes with limited forward-facing camera movement, such as the bunch of fruits, as well as scenes with 360-degree camera movement, such as the objects and monkey scenes. Despite the varying degrees of camera motion, the proposed algorithm reconstructs the scenes with high precision, generating coherent and visually accurate renderings. }

  \label{fig:real_gray}
  % \vspace{-3em}
\end{figure*}

\begin{figure*}[h!]
  \centering
  \includegraphics[width=0.9\textwidth]{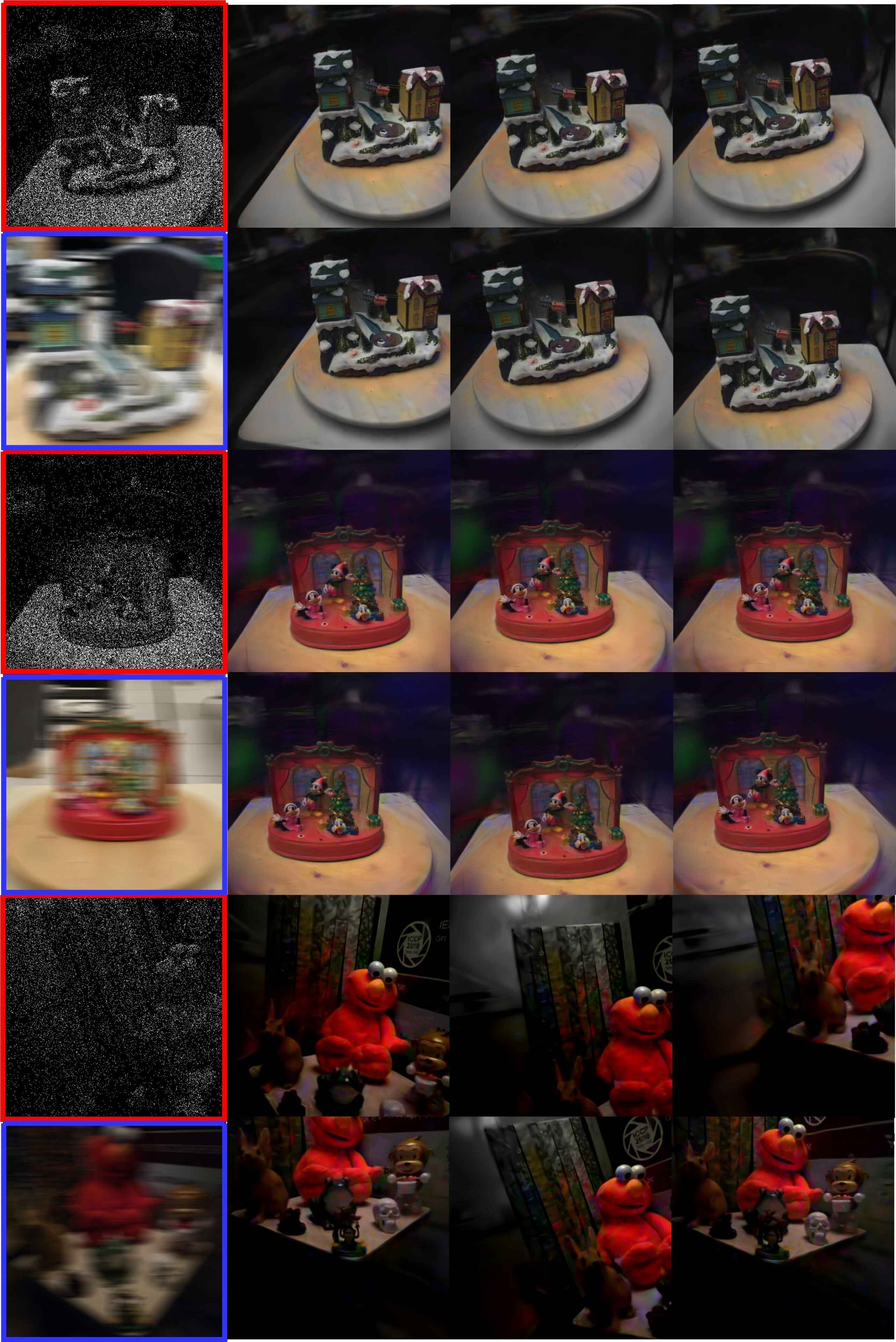}
  \caption{Qualitative Results on more real-world scenes for reference-based colorization task. The first column includes the binary image alongside its corresponding reference motion-blurred image of the real scene. The remaining columns display a series of novel view renders. Although the input image contains extreme motion blur, PhotonSplat produces clear and detailed reconstructions, maintaining consistency and accurately preserving scene features without any blurring.}
  \label{fig:real_images_renders_blurred}
  % \vspace{-3em}
\end{figure*}

\begin{figure*}[h!]
  \includegraphics[width=\textwidth]{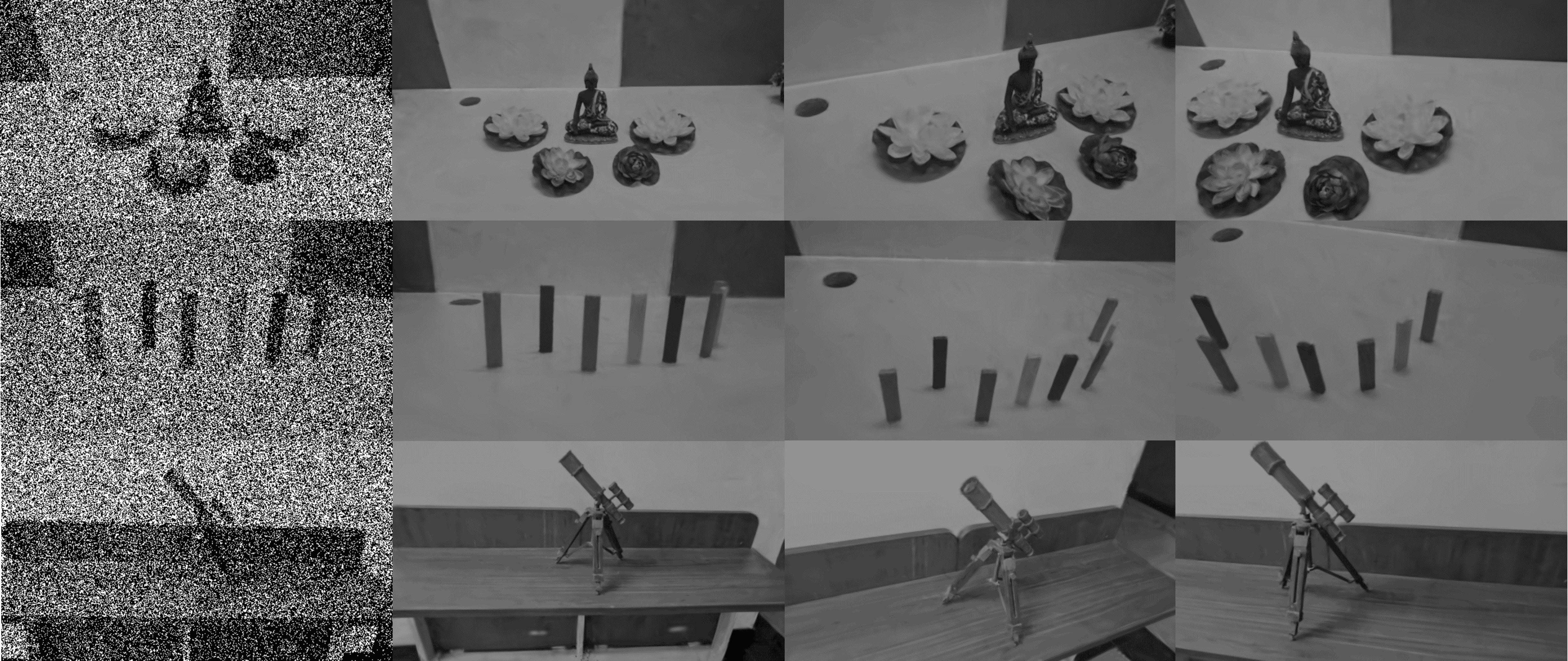}
\caption{3D reconstruction results are shown for three synthetic scenes. The first column displays the source binary image, while the subsequent three columns showcase novel-rendered views generated using PhotonSplat. The results highlight the effectiveness of our algorithm in capturing fine details from multi-view binary data and reconstructing accurate and consistent views.
}

  \label{fig:synthetic_results}
  % \vspace{-3em}
\end{figure*}

\begin{figure*}[h!]
  \includegraphics[width=\textwidth]{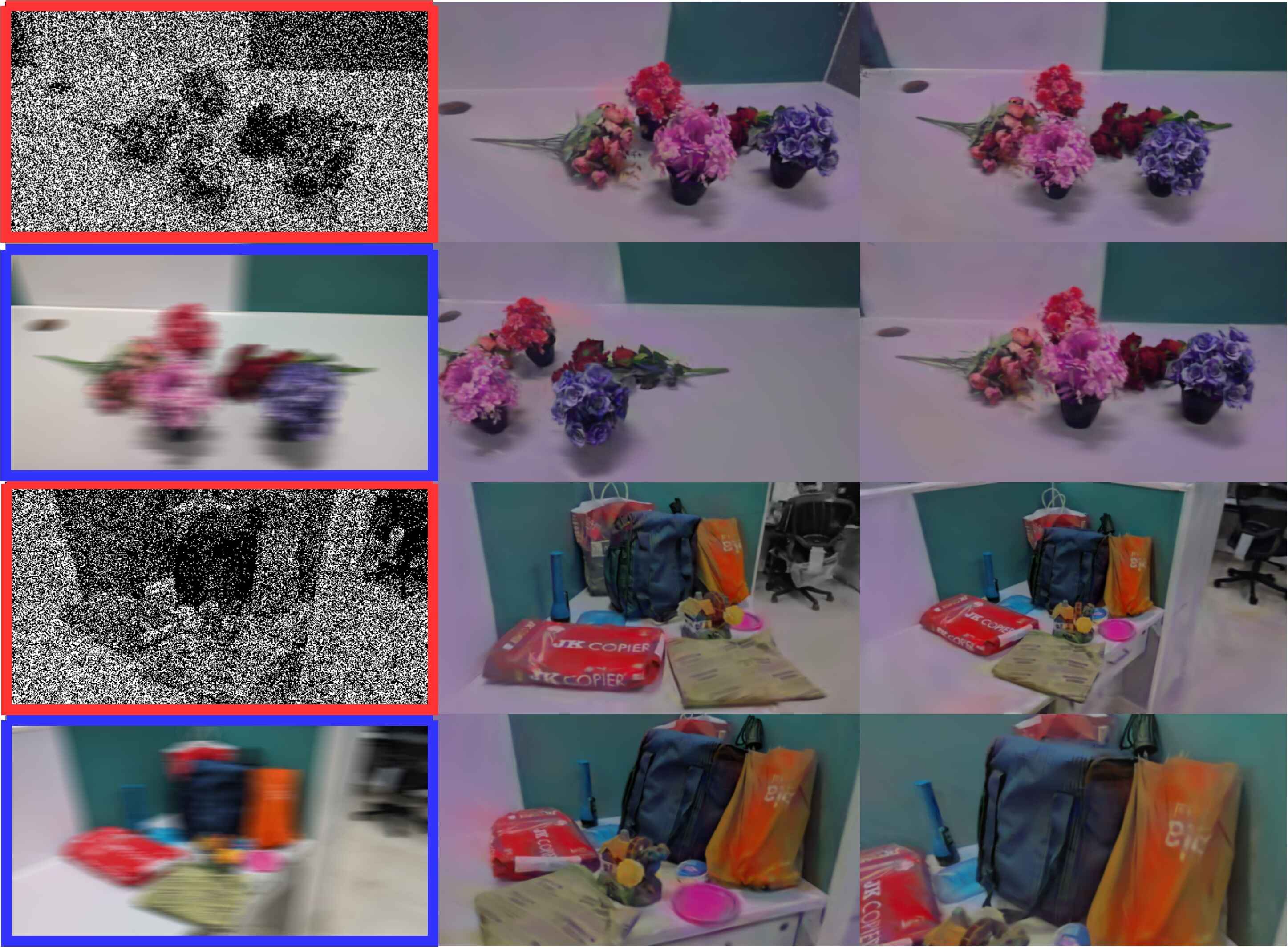}
\caption{We present qualitative results for reference-based colorization on the synthetic dataset. The first column contains the input images, where the red-highlighted regions correspond to the single photon image, and the blue box indicates the reference motion-blurred image. The next two columns form a 2×2 grid showcasing the colorized novel view renderings generated using PhotonSplat. Despite the motion blur in the input image, the outputs from PhotonSplat exhibit consistent and high-fidelity reconstructions, effectively preserving scene details without introducing any blur.
}

  \label{fig:synthetic_colo}
  % \vspace{-3em}
\end{figure*}

In Figure~\ref{fig:real_gray} and Figure~\ref{fig:synthetic_results}, we provide additional examples of reconstruction outputs using binary images as input. These results include both real-world and synthetic datasets, demonstrating the robustness of our method across diverse scenes featuring various object types and backgrounds. Additionally, we present reference-based colorization outputs in Figure~\ref{fig:real_images_renders_blurred} and Figure~\ref{fig:synthetic_colo}, evaluated on both real-world and synthetic datasets. Multi-view results across different scenes are included to highlight the view consistency achieved by our approach.

\section{Our Dataset: PhotonScenes}

\begin{figure*}[h!]
  \includegraphics[width=\textwidth]{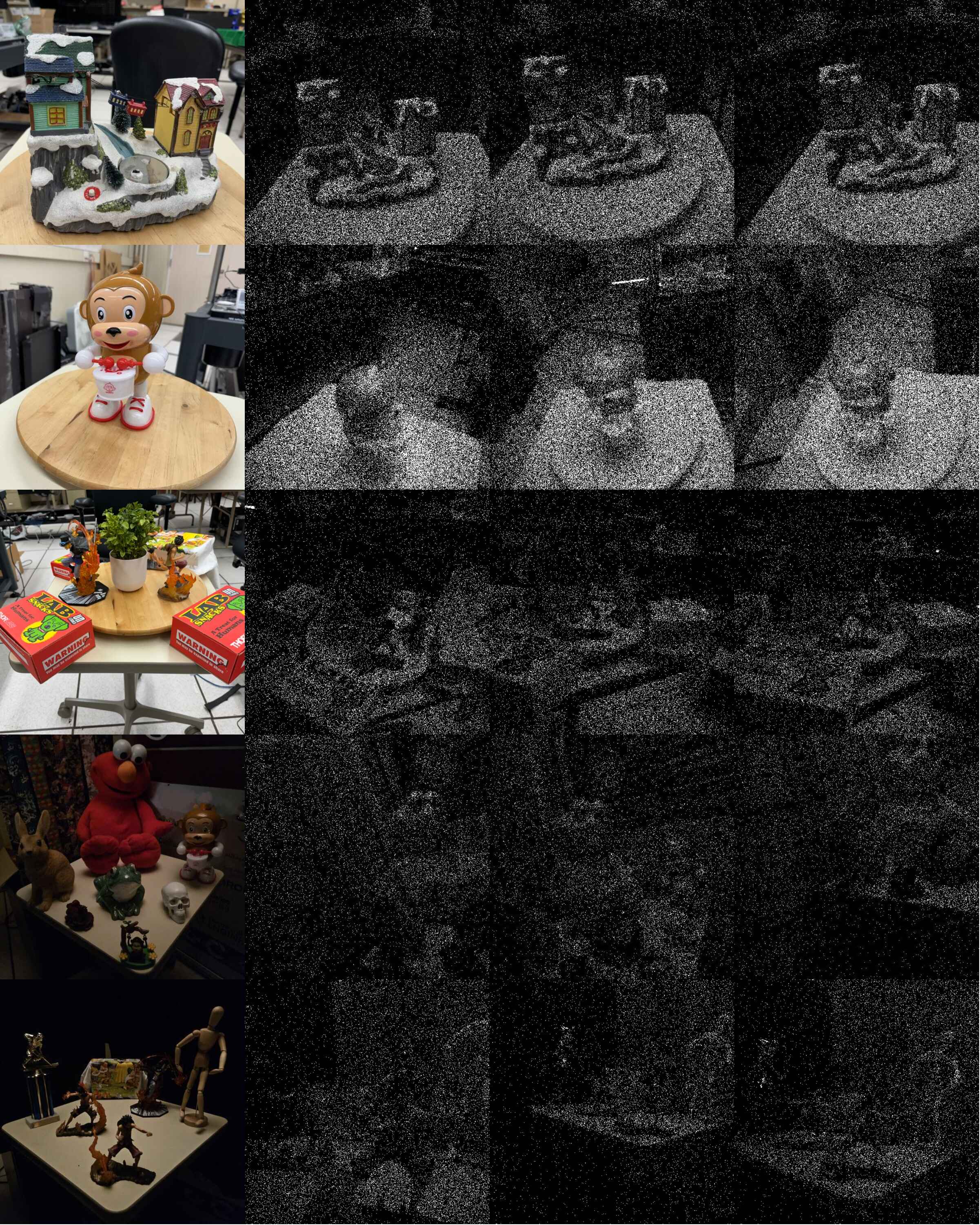}
\caption{PhotonScenes Dataset Overview: The dataset consists of multi-view SPC images alongside a single color image, as illustrated in the figure. It includes scenes captured under varying lighting conditions and camera motions to evaluate the robustness of the proposed methods. The snow-house and monkey scenes, characterized by forward-facing and 360-degree camera motion, respectively, feature high-lighting conditions. In contrast, the plant scene exhibits low lighting and moderate camera motion. Notably, the teddy bear and toys scenes demonstrate low lighting accompanied by dramatic camera motion, as reflected in the sparse white pixels of the SPC images and the corresponding RGB images. This diversity in lighting conditions and camera motions ensures the dataset's variability, offering a comprehensive evaluation framework for the proposed approach.
}

  \label{fig:dataset_picture_blurred}
  % \vspace{-3em}
\end{figure*}

Figure~\ref{fig:dataset_picture_blurred} presents a selection of ground truth views from our PhotonScenes dataset. This dataset comprises 9 real-world scenes featuring diverse objects, backgrounds, and lighting conditions. We also display several binary frames alongside an averaged grayscale image of each scene.

% \begin{figure}[h!]
%   \includegraphics[width=0.48\textwidth]{figures/main_failure_case_cropped.png}
% \caption{Our method fails to reconstruct images in extreme low-light due to insufficient information in the input SPAD images.}
%   \label{fig:fail}
%   % \vspace{-3em}
% \end{figure}

\section{Denoising Model for No-Reference Colorization}

Artifacts introduced during Gaussian splitting renders, such as oval- and circle-shaped regions with varying opacity, can significantly degrade image quality and hinder the performance of downstream image processing models. Pre-trained image colorization models, in particular, are sensitive to input quality and fail when processing degraded images. To address this, denoising is performed as a critical pre-processing step to restore clean, artifact-free images and ensure high-quality inputs for subsequent models.

A U-Net-based architecture, Uformer, was employed for this denoising task. Designed specifically for image restoration, Uformer combines a hierarchical encoder-decoder architecture with Locally-Enhanced Window (LeWin) Transformer blocks. These blocks leverage window-based self-attention and depth-wise convolutional layers to efficiently capture both local details and global dependencies while reducing computational complexity. Additionally, its Multi-Scale Restoration Modulator refines features across decoder layers, restoring fine image details. Trained on synthetic images with artificially introduced artifacts using the Flickr30k dataset and synthetic renders, Uformer effectively removes degradations. This ensures that subsequent models, like image colorization, receive high-quality inputs, significantly improving the overall pipeline performance.

For Uformer and U-Net in denoising tasks, a fixed set of hyperparameters that works well includes a learning rate of \(1 \times 10^{-4}\) and a batch size of 16. The models are trained for 100 epochs to ensure convergence, with a dropout rate of 0.3 to mitigate overfitting. The Adam optimizer is employed for its balance of convergence speed and stability, while Mean Squared Error (MSE) serves as the loss function to minimize the difference between the predicted and ground truth images along with the GAN Loss. Input image dimensions are set to \(512 \times 512\), providing a balance between computational efficiency and image detail preservation.

% \section{Additional Limitations}

% Despite its strong performance on real-world and synthetic datasets, our framework has several other limitations. Firstly, in our framework, we use 1,000 to 4,000 frames for 3D reconstruction and colorization, in contrast to the 100k frames captured by SPAD sensors. For scenes with extremely fast motion, a higher number of frames will need to be integrated. Since our framework relies on COLMAP for Structure from Motion (SfM), its inability to handle larger numbers of frames becomes a bottleneck. Future work could explore spline-based pose interpolation to generate intermediate poses and extend our framework to handle higher frames. Secondly, for reference image-based colorization, we first deblur the blurry images and then run SfM. Future works could build on integrating pose information from the IMU chip to remove the dependency on SfM and preprocessing in real-time applications.

\end{document}